\documentclass[journal]{IEEEtran}

\usepackage{xcolor,soul,framed} %,caption

\colorlet{shadecolor}{yellow}
\usepackage[pdftex]{graphicx}
\graphicspath{{../pdf/}{../jpeg/}}
\DeclareGraphicsExtensions{.pdf,.jpeg,.png}
\usepackage{subfigure}
\usepackage[cmex10]{amsmath}
\usepackage{array}
\usepackage{mdwmath}
\usepackage{mdwtab}
\usepackage{eqparbox}
\usepackage{url}
\usepackage{amssymb}
\usepackage{cite}
\usepackage{algorithm}
\usepackage{algpseudocode}

\usepackage{subfig}
\usepackage{color}
\usepackage{graphicx}%2
\usepackage{float}% 图片浮动位置
\usepackage{caption}
\begin{document}

    \title{Joint Radar Sensing, Location, and Communication Resources Optimization in 6G Network}
  \author{Haijun Zhang,~\IEEEmembership{Fellow,~IEEE,}
      Bowen Chen, Xiangnan Liu, Chao Ren
% <-this % stops a space

  \thanks{This work was supported in part by the National Natural Science Foundation of China under Grant 62225103, Grant U22B2003, and Grant 62341103; in part by the Beijing Natural Science Foundation under Grant L212004; in part by the Xiaomi Fund of Young Scholar, and in part by the National Key Laboratory of Wireless Communications Foundation under Grant IFN20230201. (\emph{Corresponding author: Haijun Zhang}.)
  
  Haijun Zhang, Bowen Chen, Xiangnan Liu, and Chao Ren are with Beijing Engineering and Technology Research Center for Convergence Networks and Ubiquitous Services, University of Science and Technology Beijing, Beijing 10083, China, (email: zhanghaijun@ustb.edu.cn; M202220848@xs.ustb.edu.cn; xiangnan.liu@xs.ustb.edu.cn; chaoren@ustb.edu.cn).}
}

\markboth{
}{Roberg \MakeLowercase{\textit{et al.}}: High-Efficiency Diode and Transistor Rectifiers}

% ====================================================================
\maketitle

% === ABSTRACT ==================================================================
\begin{abstract}
%\boldmath
The possibility of jointly optimizing location sensing and communication resources, facilitated by the existence of communication and sensing spectrum sharing, is what promotes the system performance to a higher level. However, the rapid mobility of user equipment (UE) can result in inaccurate location estimation, which can severely degrade system performance. Therefore, the precise UE location sensing and resource allocation issues are investigated in a spectrum sharing sixth generation network. An approach is proposed for joint subcarrier and power optimization based on UE location sensing, aiming to minimize system energy consumption. The joint allocation process is separated into two key phases of operation. In the radar location sensing phase, the multipath interference and Doppler effects are considered simultaneously, and the issues of UE's location and channel state estimation are transformed into a convex optimization problem, which is then solved through gradient descent. In the communication phase, a subcarrier allocation method based on subcarrier weights is proposed. To further minimize system energy consumption, a joint subcarrier and power allocation method is introduced, resolved via the Lagrange multiplier method for the non-convex resource allocation problem. Simulation analysis results indicate that the location sensing algorithm exhibits a prominent improvement in accuracy compared to benchmark algorithms. Simultaneously, the proposed resource allocation scheme also demonstrates a substantial enhancement in performance relative to baseline schemes.
\end{abstract}

% === KEYWORDS 

% === I. INTRODUCTION =============================================================
% =================================================================================
\section{Introduction}

\IEEEPARstart{W}{ith} the evolution of the sixth generation (6G) communication networks, the integrated of radar sensing, location, and communication (IRSLC) has emerged as a pivotal research field \cite{r1}. Specifically, the fusion of sensing and location with traditional communication technologies offers insights into managing complex network environments more effectively \cite{r2}. Additionally, the joint allocation of subcarriers and power aims to optimize resource allocation, thereby enhancing communication performance and energy efficiency \cite{r3}. However, the integration of sensing and location technology with subcarrier and power allocation techniques brings several challenges, primarily the effective fusion of these technologies, requiring in-depth research. Furthermore, developing an algorithm that can simultaneously ensure efficient communication and high-precision location is a current research focus \cite{r4}. Thus, the exploration of integrated network combining sensing and location with subcarrier and power joint allocation, which lays the groundwork for more intelligent and perceptive communication network \cite{r5,r6}.

Recent studies have concentrated on wireless sensing and location in communication and radar systems. During the localization process of user equipment (UE), challenges such as multipath propagation interference, random channel fading, and Doppler effects frequently emerge. Some researchers considered the Doppler effect caused by target mobility in UE position estimation. For example, Reference \cite{r7} used millimeter-wave radar to estimate the location of moving UEs, combining distance Doppler processing with digital beamforming for UE location. Reference \cite{r8} studied the maximum likelihood algorithm, constructed a Doppler channel fading model for uniform antenna arrays (ULA), and jointly estimated the Doppler shift and channel state. References \cite{r9} and \cite{r10}, focusing on random fading, multiple indirect reflect paths interference, and complex coupling channel system models, established closed-form Cramer-Rao lower bounds for target sensing. They used echoes from downlink multicarrier communication signals to jointly deduce target positions and channel states. Some researchers have used auxiliary devices to enhance target estimation accuracy. Reference \cite{r11} introduced reconfigurable distributed antennas and reflective surfaces to assist uplink signal transmission and sensing, achieving reliable UE location without compromising communication rate. Reference \cite{r12} established an integrated of sensing and communication system, supported by a intelligent reflective surface (IRS), and proposed a total least squares estimating signal parameter via rotational invariance techniques (TLS-ESPRIT) location estimation scheme. This scheme achieves precise localization even with a limited quantity of semi-passive reflective units and shorter time of sensing. Moreover, using channel parameters such as time of arrival \cite{r13,r14}, time difference of arrival \cite{r15,r16}, angle of arrival (AoA) \cite{r17,r18}, and received signal strength \cite{r19,r20} to estimate UE positions has also been a focal point for researchers. In response to the large estimation errors in AoA caused by the misidentification of multipath components, an AoA measurement method is proposed in \cite{r21}, which effectively improves positioning performance. In Reference \cite{r22}, the extraction of AoA and time difference of arrival between UE and related subarrays in ultra large multiple input multiple output channels, which is used to obtain the active coordinates of UE. However, these methods are challenging to apply directly within IRSLC and may lead to significant positioning errors \cite{r23}. 

Besides radar sensing for UE location, researchers have also concentrated on the allocation of subcarrier and power resources. A collaborative allocation method of power and bandwidth is proposed by References \cite{r24} and \cite{r25}, of which the sum of Cramer-Rao lower bounds for distance and azimuth estimation is minimized, subject to achievable communication and rate constraints. To enhance radar and communication performance, Reference \cite{r3} proposed an integrated of sensing and communication system according to subcarrier multiplexed orthogonal frequency division multiplexing (OFDM). Reference \cite{r26} investigated power allocation in cluttered environments, considering the coexistence of radar and multicarrier waveform communication systems and designing unilateral and joint power allocation schemes. In order to optimize energy in the radar and communication system, Reference \cite{r27} addressed spectral uncertainty of the target and proposed reliable radar waveform design method, which minimize power while considering for practical energy, interference, or complete ignorance at the radar receiver. References \cite{r28} and \cite{r29} introduced a method that optimizes the subcarrier and transmission power separately assigned to location sensing and communication, achieving energy savings by minimizing the system's total power.

However, most of the aforementioned works separate the design of UE sensing location and resource allocation, with few joint design schemes. An IRSLC system with distributed semi-passive IRS is proposed in Reference \cite{r30}, including a data transmission protocol for communication and sensing processes, and optimized beamforming on the basis of estimating UE positions. However in the context of IRS-assisted IRSLC system, sensing location and data transmission cannot occur simultaneously (i.e., they do not share the same time resources), making it less of a true IRSLC system. Reference \cite{r31} considered the rapid time-varying characteristics of wireless channels for maximizing the energy efficiency, which makes channel resource allocation a challenging problem, and proposed a location information-based channel resource allocation scheme for communication. While the mentioned schemes focus primarily on the allocation of communication channel resources, the aim is to design a system that combines UE sensing location,  joint optimization of subcarrier and power resources, and achieves precise UE target location and system energy consumption reduction.

In this paper, a joint sensing location and resource allocation method within an IRSLC system is proposed, where location sensing and resource allocation are putting occur concurrently, sharing equivalent time and frequency resources. The contributions can be summarized as follows:

\begin{itemize}
    \item Establishment of an IRSLC system where the base station (BS) emits downlink communication signals for location sensing and communication, capturing UE positions. The acquired location information is then utilized for joint allocation of subcarrier and power, aiming to reduce the system energy consumption and enhance its energy-saving performance.

\end{itemize}
\begin{itemize}
    \item Introduction of a positioning scheme that addresses challenges such as multipath reflection interference, random channel fading, and Doppler effect encountered when the BS perceives mobile UE locations. The UE's location and channel state estimation issues is converted to a convex problem, which tackles with gradient descent.

\end{itemize}
\begin{itemize}
    \item Formulation of a joint subcarrier and power allocation scheme where power allocation is designed based on UE location sensing information, while subcarrier allocation is dynamically assigned, taking into account the signal-to-interference plus noise ratio (SINR) for radar location sensing and communication.
\end{itemize}

The rest of paper is organized as follows: In Section II, the IRSLC system model is presented. Section III introduces the suggested location sensing scheme. Based on the UEs’ location sensing information, a joint subcarrier and power allocation scheme is designed in Section IV. Section V provides numerical results and discussions. Finally, we conclude in Section VI.

\section{SYSTEM MODEL}

% =======
% FIG. 01
% =======
\begin{figure}
  \begin{center}
  \includegraphics[width=1\linewidth]{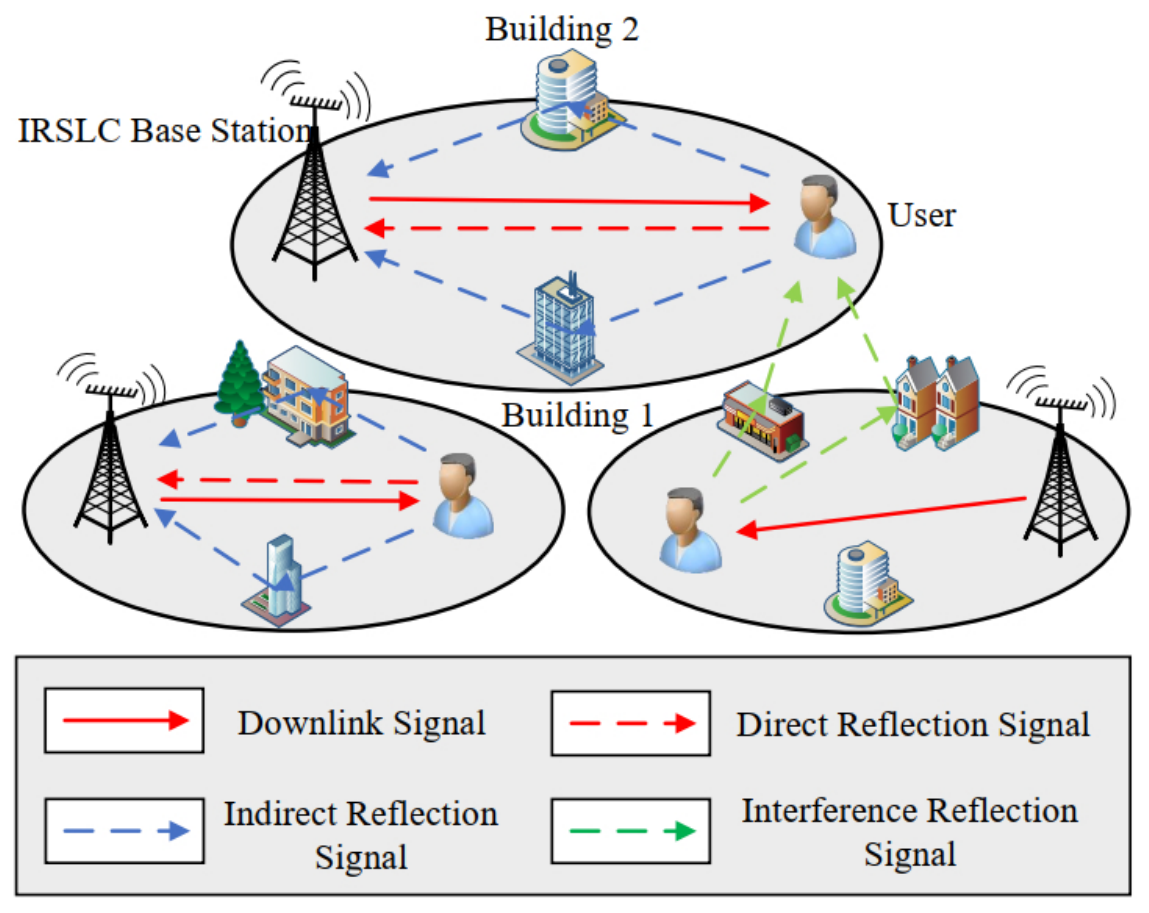}\\
  \caption{Illustration of IRSLC system model.}\label{circuit_diagram}
  \end{center}
\end{figure}

In Fig. 1, 6G IRSLC network is constructed in multiple cells, wherein the IRSLC BS is deployed with $N_T$ transmit antennas and $N_R$ receive antennas, arranged as a ULA in each cell. In the downlink of a cell, the IRSLC BS transmits OFDM signals to a moving UE within the communication region, while estimating the UE's position through echo signals reflected by the UE. The number of subcarriers in the coherent bandwidth is denoted by $N_C$. The system operates in two stages. Firstly, during the location sensing phase, the system transmits sensing signals to sense the UE's status, obtains the UE's location and channel status, and then proceeds to the communication. In the communication phase, the system continuously transmits the communication signal while also transmitting the sensing signal for information exchange between the UE and the IRSLC BS. By using downlink echo signals for sensing UE states, signaling overhead is reduced, allowing the uplink communication link to be dedicated to data transmission, enhancing overall transmission efficiency. In this paper, several assumptions are made: the lengths of OFDM signals and cyclic prefixes exceed the maximum; within the channel's coherence time, round-trip delay time of flight, angle of departure (DoA), AoA, and channel fading coefficients of the echoed signals remain constant. Additionally, the location and horizontal angle of antenna array are supposed to be known. Let ${{\bf{Q}}_{BS}} = \left( {{x_{BS}},{y_{BS}}} \right)$ denote the location of the BS antenna array, ${{\bf{Q}}_{UE}} = \left( {{x_{UE}},{y_{UE}}} \right)$ represent the unknown location of the UE, and $v_{UE}$ indicate the radial velocity of the UE. Given that the time interval is relatively small, it is assumed that there is no significant change in the UE’s velocity.

\subsection {Sensing Channel Model}

Considering a multipath reflection propagation scenario, the IRSLC BS transmits OFDM signals to the UE, including direct reflection (DR) signals and indirect reflection (IDR) interference signals reflected from other buildings, with a path count denoted as $K$. Taking the IRSLC BS antenna array as a reference, $\phi$ represents the transmission angle of the downlink signal, and ${\theta _k}$ represents the AoA of the $k$-th reflection path of the downlink. ${\tau _k}$ and ${\alpha _k}$ respectively represent the round-trip propagation delay and channel fading parameter of the $k$-th reflection paths. Here, $k = 0$ represents the DR signal, and $k= 1:K$ denotes the $k$-th IDR signals. According to the channel reciprocity principle, the AoA ${\theta _0}$ of the DR signal is equal to the DoA $\phi$ of the downlink signal. Since all reflection signals originate from the downlink signal, the downlink signal DoA of the IDR signals is also $\phi$ . Let ${f_d}$ indicate the Doppler shift caused by UE movement, which is related to the radial velocity $v_{UE}$ by ${f_d} = 2{v_{UE}}{f_c}/c$, where ${f_c}$ is the carrier frequency and $c$ is the speed of light. For DR signal, the propagation parameters ${\tau _0}$, AoA ${\theta _0}$, and the DoA $\phi$ are related to the UE location ${{\bf{Q}}_{UE}}$ and satisfy ${\tau _0} = 2{\left\| {{{\bf{Q}}_{UE}} - {{\bf{Q}}_{BS}}} \right\|_2}/c$ and ${\theta _0} = \phi  = \arccos \left( {\left( {{{\bf{Q}}_{UE}} - {{\bf{Q}}_{BS}}} \right){{\bf{e}}_x}/{{\left\| {{{\bf{Q}}_{UE}} - {{\bf{Q}}_{BS}}} \right\|}_2}} \right)$, where ${{\bf{e}}_x} = {[1,0]^T}$. After giving the above propagation parameters, the downlink channel state matrix of the $k$-th reflection path on the $n$-th subcarrier is $\mathbf{H}_{rad,n,k}\in\mathbb{C}^{N_R\times N_T}$, which represents the transmission coefficient from the transmitter, can be expressed as \cite{r9}
\begin{equation}
    \mathbf{H}_{rad,n,k}=\alpha_ke^{-j2\pi(f_n-f_d)\tau_k}\mathbf{b}_n(\theta_k)\mathbf{a}_n^H(\phi),
\end{equation}
where \({f_n}\) represents the baseband frequency of the $n$-th subcarrier, satisfying
\begin{equation}
f_n=\frac n{NT_s},n=1:N_c,
\end{equation}
where $N$ is the number of system carriers, and $T_s$ is the sampling period, remaining constant within the coherent bandwidth. This is equivalent to the system bandwidth being separated into multiple sub-bands by coherent bandwidth and using the OFDM signal within a sub-band to estimate channel. In addition, ${{\bf{b}}_n}\left( {{\theta _k}} \right) \in {{\mathbb C}^{{N_R} \times 1}}$ and ${{\bf{a}}_n}\left( \phi  \right) \in {{\mathbb C}^{{N_T} \times 1}}$ respectively represent the receiving and transmitting steering vectors of the IRSLC BS, which are given by \cite{r32}
\begin{equation}
   \mathbf{a}_n(\phi)=\left[1,\ldots,e^{-j2\pi(t-1)f_n^{^{\prime}}d_A\cos\phi}\right]^T,t=1:N_T,
\end{equation}
\begin{equation}
   \mathbf{b}_n(\theta_k)=\left[1,\ldots,e^{-j2\pi(r-1)f_n^{^{\prime}}d_A\cos\theta_k}\right]^T,r=1:N_R,
\end{equation}where $f_n^{\prime}=f_n+f_c$ represents the subcarrier frequency, $d_A$ represents the element spacing, equal to half the wavelength of the carrier signal.

\subsection {Communication Channel Model}
The location of UE is randomly distributed within the cell, the UE may undergoes varying channel gains. The UE's path loss is constructed as follows \cite{r33}\begin{equation}
    PL=10\log_{10}(\frac{4\pi f_cd_0}c)^2+10\log_{10}(\frac{d_{U2B}}{d_0})^\varepsilon+\zeta,
\end{equation}where $d_0$ and $\varepsilon$ represent the reference distance and path loss exponent, respectively. Furthermore, $d_{U2B}=\left\|\mathbf{Q}_{UE}-\mathbf{Q}_{BS}\right\|_2$ indicates the distance from the UE to the IRSLC BS, and $\zeta$ represents the log-normal shadowing fading, which is a Gaussian random variable with a mean of $0$ and a variance of $\sigma_{\zeta}^2$. Consequently, the large-scale fading for the UE is given by $\beta=10^{-PL/10}$. Given the aforementioned channel propagation parameters, the communication channel state matrix for the downlink of the UE on the $n$-th subcarrier is
\begin{equation}
\mathbf{H}_{com,n}=\sqrt{\beta}\mathbf{a}_n^H\left(\phi\right)\in\mathbb{C}^{N_T\times1}.
\end{equation}

\subsection {Signal Model}

During the periods of location sensing phase, the IRSLC BS transmits OFDM symbols to sense the UE. The radar echo signal  $\mathbf{y}_{rad,n}\in\mathbb{C}^{N_R\times1}$ is shown as
\begin{equation}
\mathbf{y}_{rad,n}=\mathbf{H}_{rad,n,0}\mathbf{W_R}\mathbf{s}_m\left(n\right)+\sum_{k=1}^K\mathbf{H}_{rad,n,k}\mathbf{W_R}\mathbf{s}_m\left(n\right)+\mathbf{n}_R,
\end{equation}
where the first part corresponds to the echo signal from DR, and the second part accounts for the interfering echo signals from the $K$ IDR; $\mathbf{W_R}\in\mathbb{C}^{N_T\times{N_T}}$ represents the sensing beamforming matrix; $\mathbf{s}_m\left(n\right)\in\mathbb{C}^{N_T\times1}$ represents the $m$-th OFDM symbol transmitted on the $n$-th sub-carrier, with $\mathbb{E}\left[\left|\mathbf{s}_m\right|^2\right]=1$; $\mathbf{H}_{rad,n,0}$ denotes the channel matrix for the DR component, and $\mathbf{H}_{rad,n,k}$ represents the channel matrix for the IDR component; $\mathbf{n}_R\in\mathbb{C}^{N_R\times1}$ is the additive white Gaussian noise (AWGN) matrix, with a variance of $\sigma_{R}^{2}$. 

Therefore, by substituting (1) into (7), we can derive the explicit expression for the echo signal as follows
\begin{equation}
    \begin{aligned}
\mathbf{y}_{rad,n}& =\alpha_{0}e^{-j2\pi(f_{n}-f_{d})\tau_{0}}\mathbf{b}_{n}(\theta_{0})\mathbf{a}_{n}^{H}(\phi)\mathbf{W_R}\mathbf{s}_m\left(n\right)+  \\
&\sum_{k=1}^K\alpha_ke^{-j2\pi(f_n-f_d)\tau_k}\mathbf{b}_n(\theta_k)\mathbf{a}_n^H(\phi)\mathbf{W_R}\mathbf{s}_m\left(n\right)+\mathbf{n}_R. 
\end{aligned}
\end{equation}

In the sensing phase, the sensing SINR under the $n$-th subcarrier as follows \cite{r34}
\begin{equation}
    SINR_{rad,n}=\frac{\alpha_0\left|\mathbf{b}_n\left(\theta_0\right)\mathbf{a}_n^H\left(\phi\right)\right|^2\left|w_R\right|^2}{\sum_{k=1}^K\alpha_k\left|\mathbf{b}_n\left(\theta_k\right)\mathbf{a}_n^H\left(\phi\right)\right|^2\left|w_R\right|^2+\sigma_R^2}.
\end{equation}

In the communication phase, the signal received by the communication receiver from IRSLC BS transmission can be expressed as
\begin{equation}
    \begin{aligned}y_{com,n}&=\mathbf{H}_{com,n}\mathbf{W_C}\mathbf{s}_m\left(n\right)+\\&\sum_{k=1}^{K}\alpha_ke^{-j2\pi(f_n-f_d)\tau_k}b_n\left(\theta_k\right)\mathbf{a}_n^H\left(\phi\right)\mathbf{W_R}\mathbf{s}_m\left(n\right)+\mathbf{n}_c,\end{aligned}
\end{equation}
where the first item is the signal transmitted by the IRSLC BS, the second item is the radar sensing interference signal from other cells, and the third item is the communication with $\sigma_{C}^{2}$. $\mathbf{W_C}\in\mathbb{C}^{N_T\times{N_T}}$ represents the communication beamforming matrix. Therefore, considering Eq. (6), the receiving SINR of the UE under the $n$-th subcarrier is
\begin{equation}
    SINR_{com,n}=\frac{\beta\left|\mathbf{a}_{n}^{H}\left(\phi\right)\right|^{2}\left|w_{C}\right|^{2}}{\sigma_{c}^{2}+\sum_{k=0}^{K}\alpha_{k}\left(b_{n}\left(\theta_{k}\right)\right)^{2}\left|\mathbf{a}_{n}^{H}\left(\phi\right)\right|^{2}\left|w_{R}\right|^{2}}.
\end{equation}

\section{LOCATION SENSING DESIGN}

\subsection {Problem Formulation}

In the UE location sensing phase, to acquire UE location and channel state information, let $\mathbf{U}=\left[\mathbf{Q}_{UE},v_{UE}\right]$ represent the UE’s location and velocity information, $\mathbf{H}=\left[\mathbf{H}_{rad,n,0},\ldots,\mathbf{H}_{rad,n,L}\right]$ denote the channel state information of the sensing channel, $\mathbf{\rho}=\left\{\theta_{k},\tau_{k}\left|\forall k=1\colon K\right\}\right.$ signify the reflection propagation parameters of IDR, and $\mathbf{y}_{rad}\in\mathbb{C}^{N_{R}N_{C}M\times1}$ represent the received signals at $N_R$ receiving antennas, which transmit $M$ OFDM symbols under $N_c$ sub-carriers, given by
\begin{equation}
    \mathbf{y}_{rad}=\left[\mathbf{y}_{rad,1,1},\ldots,\mathbf{y}_{rad,1,M},\ldots,\mathbf{y}_{rad,n,M},\ldots,\mathbf{y}_{rad,N,M}\right]^T.
\end{equation}

Let $\mathbf{f}_{DR}(\mathbf{H},\mathbf{U})\in\mathbb{C}^{N_{R}N_{C}M\times1}$ and $\mathbf{f}_{\mathit{IDR}}(\mathbf{H},\mathbf{U};\mathbf{\rho})\in\mathbb{C}^{N_{R}N_{C}M\times1}$ respectively represent the signal echo models for DR and IDR, which can be expressed as
\begin{equation}
    \mathbf{f}_{DR}\left(\mathbf{H},\mathbf{U}\right)=\left[\mathbf{f}_{DR,n,m}|\forall n\in{N}_{C},\forall m\in{M}\right],
\end{equation}
\begin{equation}
    \mathbf{f}_{IDR}\left(\mathbf{H},\mathbf{U};\mathbf{\rho}\right)=\left[\mathbf{f}_{IDR,n,m}|\forall n\in{N}_{C},\forall m\in{M}\right],
\end{equation}
wherein, the respective elements of the corresponding signal echo model are represented as
\begin{equation}
    \mathbf{f}_{DR,n,m}=\alpha_0e^{-j2\pi\left(f_n-f_d\right)\tau_0}\mathbf{b}_n\left(\theta_0\right)\mathbf{a}_n^H\left(\phi\right)\mathbf{W}_R\mathbf{s}_m[n],
\end{equation}
\begin{equation}
    \mathbf{f}_{IDR,n,m}=\sum_{k=1}^{K}\alpha_{k}e^{-j2\pi(f_{n}-f_{d})\tau_{k}}\mathbf{b}_{n}\left(\theta_{k}\right)\mathbf{a}_{n}^{H}\left(\phi\right)\mathbf{W}_{R}\mathbf{s}_{m}[n].
\end{equation}

Thus, there will be further construct the signal propagation function $\mathbf{f}\left(\mathbf{H},\mathbf{U};\mathbf{\rho}\right)=\mathbf{f}_{DR}\left(\mathbf{H},\mathbf{U}\right)+\mathbf{f}_{IDR}\left(\mathbf{H},\mathbf{U};\mathbf{\rho}\right)$. According to the signal multipath reflection model in the 6G IRSLC network, the corresponding echo signal model can be given as
\begin{equation}
    \mathbf{y}_{rad}=\mathbf{f}\left(\mathbf{H},\mathbf{U};\mathbf{\rho}\right)+\mathbf{n},
\end{equation}
where $\mathbf{n}\in\mathbb{C}^{N_{R}N_{C}M\times1}$ is a zero-mean AWGN vector. Based on this signal model, the UE location sensing design can be proposed to estimate the UE's location $\mathbf{Q}_{UE}$ and channel state information $\mathbf{H}$ using the signal echo models $\mathbf{y}_{rad}$ and $\mathbf{f}\left(\mathbf{H},\mathbf{U};\mathbf{\rho}\right)$ and the signal multipath reflection model. Consequently, the location sensing optimization can be articulated as the following minimization problem
\begin{equation}
     \hat{\mathbf{U}}=\arg\min_{\mathbf{U}}\min\left\|\mathbf{y}_{rad}-\mathbf{f}(\mathbf{H},\mathbf{U};\mathbf{\rho})\right\|_2^2.
 \end{equation}

This minimization problem faces three challenges:
\begin{enumerate}
    \item The presence of multiple unknown parameters such as $\mathbf{U}$ and $\mathbf{\rho}$ complicates the solution to the problem. 
    \item The cost function $\min\left\|\mathbf{y}_{rad}-\mathbf{f}(\mathbf{H},\mathbf{U};\mathbf{\rho})\right\|_2^2$ lacks a closed-form expression, making it hard to obtain the optimal solution for the target variable $\mathbf{U}$ even after multiple iterations. 
    \item The signal propagation function $\mathbf{f}\left(\mathbf{H},\mathbf{U};\mathbf{\rho}\right)$ is nonlinear with respect to $\mathbf{H}$, $\mathbf{U}$ and $\mathbf{\rho}$. As a result, the minimization problem posed is non-convex, necessitating a transformation for its resolution.
\end{enumerate}

\subsection {Problem Solution}

To address these challenges, the problem can be solved using the following approach: 
\begin{enumerate}
    \item The equivalent modeling is employed to incorporate complex channel parameters into the equivalent channel state matrix, reducing the difficulty in estimating IDR channel.
    \item The minimization problem is turned into a joint optimization problem by introducing a convex surrogate function for the cost function, allowing the problem to possess a closed-form expression and attain the optimal solution.
    \item The joint optimization problem is still non-convex, so it will be changed to two problems: the estimation of equivalent channel state matrix and the estimation of UE state. 

\end{enumerate}

The optimization is conducted alternately until convergence. Specifically, the complex channel parameters include the channel fading coefficient $\alpha_k$, the phase coefficient generated by propagation delay $e^{-j2\pi(f_{n}-f_{d})\tau_{k}}$, and the arrival angle gain of IDR $\mathbf{b}_{n}\left(\theta_{k}\right)$. Let $\mathbf{H}_{IDR,n}\in\mathbb{C}^{N_R}$ represent the equivalent IDR channel state matrix under the $n$-th subcarrier, which can be expressed as follows
\begin{equation}
    \mathbf{H}_{IDR,n}=\sum_{k=1}^K\alpha_ke^{-j2\pi(f_n-f_d)\tau_k}\mathbf{b}_n(\theta_k).
\end{equation}

Let ${\mathbf{H}}_{E{\mathcal{Q}}}=\left[H_{DR};{\mathbf{H}}_{IDR}\right]\in\mathbb{C}^{(N_{R}N_{C}+1)\times1}$ denote the equivalent channel state matrix for the process in which the sensing signal is transmitted and reflected back to BS, with $\mathbf{H}_{IDR}\in\mathbb{C}^{N_{R}N_{c}\times1}$. According to Eq. (17), the reflected signal $\mathbf{y}_{rad}$ can be equivalently represented as
\begin{equation}
    \mathbf{y}_{rad}=\mathbf{F}(\mathbf{U})\mathbf{H}_{E\mathcal{Q}}+\mathbf{n},
\end{equation}
where $\mathbf{F}(\mathbf{U})\in\mathbb{C}^{N_{R}N_{c}M\times(N_{R}N_{C}+1)}$ is the coefficient matrix of $\mathbf{H}_{E\mathcal{Q}}$, and it depends on the UE state $\mathbf{U}$, expressed as $\mathbf{F}(\mathbf{U})=\begin{bmatrix}\mathbf{G}(\mathbf{U}),\mathbf{F}(\mathbf{Q}_{UE})\end{bmatrix}$. For the first part of the coefficient matrix $\mathbf{G}(\mathbf{U})=\left[\mathbf{G}_{n,m}(\mathbf{U})|\forall n\in{N}_{C},\forall m\in{M}\right]\in\mathbb{C}^{N_{R}N_{C}M\times1}$, the detailed expression for the first part of the coefficient matrix is
\begin{equation}
    \mathbf{G}_{n,m}\left(\mathbf{U}\right)=e^{-j2\pi(f_n-f_d)\tau_0}\mathbf{b}_n\left(\mathbf{Q}_{UE}\right)\mathbf{a}_n^H\left(\mathbf{Q}_{UE}\right)\mathbf{W}_R\mathbf{s}_m(n).
\end{equation}

For the second part of the coefficient matrix $\mathbf{F}(\mathbf{Q}_{UE})=\begin{bmatrix}\mathbf{F}_m(\mathbf{Q}_{UE})|\forall m\in {M}\end{bmatrix}\in\mathbb{C}^{N_{R}N_{\boldsymbol{C}}M\times N_{R}N_{\boldsymbol{C}}}$, its specific expression is
\begin{equation}
    \mathbf{F}_{n,m}\left(\mathbf{Q}_{UE}\right)=\mathbf{a}_{n}^{H}\left(\mathbf{Q}_{UE}\right)\mathbf{W}_{R}\mathbf{s}_{m}(n)\mathbf{I}_{N_{R}}\in\mathbb{C}^{N_{R}\times N_{R}},
\end{equation}
where $\left.\mathbf{F}_m\left(\mathbf{Q}_{UE}\right)=dia\mathbf{g}\left[\mathbf{F}_{n,m}\left(\mathbf{Q}_{UE}\right)\right|\forall n\in{N}_{C}\right]$, and $\mathbf{I}_{N_{R}}$ is a $N_{_R}\times N_{_R}$ identity matrix. In the aforementioned transformation, all unknown parameters of IDR are incorporated into the equivalent channel state matrix $\mathbf{H}_{E\mathcal{Q}}$, effectively reducing the difficulty in estimating the IDR channel, resolving issue 1).

Based on the constructed equivalent channel model, the UE location sensing issue should be further formulated as
\begin{equation}
    \hat{\mathbf{U}}=\arg\min_{\mathbf{U}}\min\left\|\mathbf{y}_{rad}-\mathbf{F}(\mathbf{U})\mathbf{H}_{EQ}\right\|_2^2.
\end{equation}

Although the minimization problem above lacks a closed-form solution, the unilateral optimization issue is transformed into a joint optimization issue to address the UE location sensing issue
\begin{equation}
    \left(\hat{\mathbf{H}}_{E\mathcal{Q}},\hat{\mathbf{U}}\right)=\arg\min_{\mathbf{H}_{E\mathcal{Q}},\mathbf{U}}\left\|\mathbf{y}_{rad}-\mathbf{F}(\mathbf{U})\mathbf{H}_{E\mathcal{Q}}\right\|_2^2.
\end{equation}

Consequently, the optimal solution could be attained by finding the equivalent channel state matrix $\mathbf{H}_{E\mathcal{Q}}$ and UE state $\mathbf{U}$. Therefore, the issue 2), which lacks a closed-form expression and cannot obtain the optimal solution, is resolved.

Since the joint optimization problem in the relation between $\mathbf{F}(\mathbf{U})$ and $\mathbf{H}_{E\mathcal{Q}}$ with respect to UE state $\mathbf{U}$ is still nonlinear. In order to solve this problem, this paper introduces a convex surrogate function related to $\left\|\mathbf{y}_{rad}-\mathbf{F}(\mathbf{U})\mathbf{H}_{EQ}\right\|_{2}^{2}$ to acquire a locally optimal solution. Moreover, since the cost function is a linear function of the equivalent channel state matrix $\mathbf{H}_{E\mathcal{Q}}$, the joint optimization problem needs to be decomposed into solving two optimization sub-problems: the estimation of equivalent channel state matrix $\mathbf{H}_{E\mathcal{Q}}$ and the estimation of UE state $\mathbf{U}$. By exploiting the convex structure of the system model, we alternately optimize the equivalent channel state matrix $\mathbf{H}_{E\mathcal{Q}}$ and UE state $\mathbf{U}$ to achieve joint optimization.

Specifically, the estimation of the equivalent channel state matrix $\mathbf{H}_{E\mathcal{Q}}$ is performed initially. Assuming that the UE state $\mathbf{U}$ is known in the $i$-th iteration and its estimated value is $\mathbf{\hat{U}}_{[i]}$, the equivalent channel state estimation sub-problem can be expressed
\begin{equation}
    \hat{\mathbf{H}}_{E\mathcal{Q}}=\arg\min_{\mathbf{H}_{E\mathcal{Q}}}\left\|\mathbf{y}_{rad}-\mathbf{F}\left(\hat{\mathbf{U}}_{[i]}\right)\mathbf{H}_{E\mathcal{Q}}\right\|_2^2.
\end{equation}

By satisfying linear structural requirements, the equivalent channel state matrix $\mathbf{H}_{E\mathcal{Q}}$ can be estimated in the current iteration as
\begin{equation}
    \hat{\mathbf{H}}_{[i]}=\left(\mathbf{F}^H\left(\hat{\mathbf{U}}_{[i]}\right)\mathbf{F}\!\left(\hat{\mathbf{U}}_{[i]}\right)\right)^{-1}\mathbf{F}^H\!\left(\hat{\mathbf{U}}_{[i]}\right)\mathbf{y}_{rad}.
\end{equation}

Given the estimated UE state value $\mathbf{\hat{U}}_{[i]}$, the optimal estimation of the equivalent channel state matrix $\mathbf{H}_{E\mathcal{Q}}$ can be obtained. Based on the optimal estimation of the equivalent channel state matrix $\hat{\mathbf{H}}_{[i]}$ obtained from the previous iteration, and according to Eq. (24), the UE state estimation is shown as
\begin{equation}
    \hat{\mathbf{U}}=\arg\min_{\mathbf{U}}\left\|\mathbf{y}_{\textit{rad}} - \mathbf{F}(\mathbf{U})\hat{\mathbf{H}}_{[i]}\right\|_2^2.
\end{equation}

To resolve the non-convex issue of UE state estimation, we introduce a convex surrogate function. Specifically, let $\Gamma\big(\mathbf{U};\hat{\mathbf{U}}_{[i]},\hat{\mathbf{H}}_{[i]}\big)$ denote the convex surrogate function of cost function $\left\|\mathbf{y}_{rad}-\mathbf{F}(\mathbf{U})\mathbf{H}_{EQ}\right\|_{2}^{2}$, which can be expressed as
\begin{equation}
    \begin{aligned}
\Gamma\Big(\mathbf{U};\hat{\mathbf{U}}_{[i]},\hat{\mathbf{H}}_{[i]}\Big)=& \parallel\mathbf{y}_{rad}-\mathbf{F}\Big(\mathbf{\hat{U}}_{[i]}\Big)\mathbf{\hat{H}}_{[i]}-  \\
&\nabla_{\mathbf{U}}^{H}\left(\mathbf{F}\Big(\mathbf{\hat{U}}_{[i]}\Big)\mathbf{\hat{H}}_{[i]}\right)\Big(\mathbf{U}-\mathbf{\hat{U}}_{[i]}\Big)\|_{2}^{2},
\end{aligned}
\end{equation}
where $\nabla_{\mathbf{U}}\left(\mathbf{F}\big(\hat{\mathbf{U}}_{[i]}\big)\hat{\mathbf{H}}_{[i]}\right)\in\mathbb{C}^{3\times N_{\mathbf{R}}N_{\mathbf{C}}M}$ is the gradient value of $\mathbf{F}(\hat{\mathbf{U}}_{[i]})\hat{\mathbf{H}}_{[i]}$ at $\mathbf{U}=\hat{\mathbf{U}}_{[i]}$, calculated as follows

\begin{equation}
    \nabla_{\mathbf{U}}\left(\mathbf{F}(\hat{\mathbf{U}}_{[i]})\hat{\mathbf{H}}_{[i]}\right)=\xi\left(\hat{\mathbf{U}}_{[i]}\right)\mathbf{\chi}_{EQ},
\end{equation}
 where $\xi\left(\hat{\mathbf{U}}_{[i]}\right)\in\mathbb{C}^{3\times(N_{R}N_{C}+1)N_{R}N_{C}M}$ and $\xi\left(\hat{\mathbf{U}}_{[i]}\right)=\left[\left(\xi_{r,n,m}\left(\hat{\mathbf{U}}_{[i]}\right)\right)^\mathrm{T}\mid\forall r\in N_\mathrm{~R},\forall n\in N_\mathrm{c},\forall m\in M\right]^\mathrm{T}$. For $\xi_{r,n,m}\left(\hat{\mathbf{U}}_{[i]}\right)$, its specific expression is as follows\begin{equation}
     \xi_{r,n,m}\left(\hat{\mathbf{U}}_{[i]}\right)=\left[\xi_{r,n,m}^{\mathbf{Q}_{UE,1}}\left(\hat{\mathbf{U}}_{[i]}\right),\xi_{r,n,m}^{\mathbf{Q}_{UE,2}}\left(\hat{\mathbf{U}}_{[i]}\right)\right],
 \end{equation}
\begin{equation}
\begin{aligned}&\xi_{r,n,m}^{\mathbf{Q}_{UE,1}}\left(\mathbf{\hat{U}}_{[i]}\right)=\\&\nabla_{\mathbf{Q}_{UE}}\left(e^{-\mathrm{j}2\pi(f_n-f_d)\hat{\tau}_{0,[i]}}b_{r,n}\mathbf{a}_{n}^{\mathrm{H}}(\mathbf{\hat{Q}}_{UE,[i]})\mathbf{W}_R\mathbf{s}_m(n)\right),\end{aligned}
\end{equation}
\begin{equation}
\xi_{r,n,m}^{\mathbf{Q}_{UE,2}}\left(\hat{\mathbf{U}}_{[i]}\right)=\nabla_{\mathbf{Q}_{UE}}\left(\mathbf{a}_{n}^{\mathrm{H}}(\mathbf{\hat{Q}}_{UE,[i]})\mathbf{W}_{R}\mathbf{s}_{m}(n)\right)\boldsymbol{\delta}_{r,n}^{\mathrm{H}},
\end{equation}
where $\hat{\tau}_{0,[i]}$ represents the round-trip propagation delay estimation, and $\mathbb{\delta}_{r,n}^{\mathrm{H}}\in\mathbb{R}^{N_{R}N_{C}\times1}$ is an indicator vector that takes a value of 1 if and only if the element index in the vector is $(n-1)N_R+r$, with all other elements being 0. The entries of $\xi_{r,n,m}^{\mathbf{Q}_{UE,1}}\left(\hat{\mathbf{U}}_{[i]}\right)$ and $\xi_{r,n,m}^{\mathbf{Q}_{UE,2}}\left(\hat{\mathbf{U}}_{[i]}\right)$ are derived in Appendix A. Furthermore, $\mathbb{\chi}_{_{EQ}}\in\mathbb{C}^{\left(N_{R}N_{C}+1\right)N_{R}N_{C}M\times N_{R}N_{C}M}$ in equation (29) can be expressed as
\begin{equation}
    \chi_{_{EQ}}=\mathbf{I}_{_{N_RN_CM}}\odot\mathbf{H}_{EQ}\left|_{\mathbf{H}_{EQ}=\hat{\mathbf{H}}_{[i]}}\right.,
\end{equation}
where $\odot$ indicates the Hadamard product.

Thus, given $\hat{\mathbf{U}}_{[i]}$, $\hat{\mathbf{H}}_{[i]}$, and $\Gamma\big(\mathbf{U};\hat{\mathbf{U}}_{[i]},\hat{\mathbf{H}}_{[i]}\big)$, the minimization problem (27) could be further expressed as
\begin{equation}    \hat{\mathbf{U}}=\arg\min_\mathbf{U}\Gamma(\mathbf{U};\hat{\mathbf{U}}_{[i]},\hat{\mathbf{H}}_{[i]}).
\end{equation}

The closed-form expression for the UE state estimation is
\begin{equation}
    \hat{\mathbf{U}}_{[i+1]}=\hat{\mathbf{U}}_{[i]}+\mathbf{\kappa}_{[i+1]},
\end{equation}
where $\mathbf{\kappa}_{[i+1]}$ denotes the update direction for $\hat{\mathbf{U}}_{[i+1]}$, and its specific expression is
\begin{equation}
    \begin{aligned}
\mathbf{\kappa}_{[i+1]}=& \left(\xi(\hat{\mathbf{U}}_{[i]})\boldsymbol{\chi}_{EQ}\big(\xi\big(\hat{\mathbf{U}}_{[i]}\big)\boldsymbol{\chi}_{EQ}\big)^H\right)^{-1}*  \\
&\left(\xi\big(\hat{\mathbf{U}}_{[i]}\big)\boldsymbol{\chi}_{E\boldsymbol{Q}}\right)\big(\mathbf{y}_{rad}-\mathbf{F}\big(\hat{\mathbf{U}}_{[i]}\big)\hat{\mathbf{H}}_{[i]}\bigg).
\end{aligned}
\end{equation}

In summary, the mobile UE location sensing estimation is presented according to the IRSLC system in Algorithm 1. Initially, the UE state value $\hat{\mathbf{U}}_{[i]}$ should be provided along with other fundamental parameters. The alternating iterative optimization is performed through Eqs. (26) and (35) until convergence. The design of Algorithm 1 is as

\begin{algorithm}
\caption{IRSLC-based UE Location Sensing}\label{alg:algorithm1}
\begin{algorithmic}[1]
\State  \textbf{Initialization} the initial state of $\hat{\mathbf{U}}_{[0]}$, transmit symbols $\mathbf{s}_m\left(n\right)$, echo signal $\mathbf{y}_{rad}$, and the number of iterations $N_{ite}$;
\State  \textbf{for} i = 1 to $N_{ite}$ \textbf{do}
\State  \quad Calculate $\hat{\mathbf{H}}_{[i]}$ by solving (26);
\State  \quad Calculate $\mathbf{\kappa}_{[i+1]}$ by solving (36);
\State  \quad Calculate $\hat{\mathbf{U}}_{[i+1]}$ by solving (35);
\State  \textbf{end for}
\State  \textbf{Output} $\mathbf{\hat{x}}_{UE}=\left(\mathbf{\hat{U}}\right)_{_{1:2}},$ $\hat{\mathbf{H}}_{EQ}=\hat{\mathbf{H}}_{[i]}$.
\end{algorithmic}
\end{algorithm}

The UE location estimation is obtained through Algorithm 1. The AoA of echo signal obtained at BS can be represented by
\begin{equation}
    \hat{\theta}_{0}=\arccos\left(\frac{\left(\mathbf{\hat{Q}}_{UE}-\mathbf{Q}_{BS}\right)^{T}e_{X}}{\left\|\mathbf{\hat{Q}}_{UE}-\mathbf{Q}_{BS}\right\|_{2}}\right).
\end{equation}

\section{POWER AND SUBCARRIER JOINT ALLOCATION DESIGN}
In this section, based on the AoA estimation obtained from the UE location sensing algorithm presented in Section III, an allocation of joint subcarrier and power scheme is designed during the communication phase to minimize system energy consumption while ensuring satisfactory performance in both the location-sensing phase and the communication phase.

\subsection {Problem Formulation}

To achieve the goals mentioned above, we introduce constraints related to the sensing quality of service (QoS) and communication QoS. The communication rate (CR) is used to evaluate communication QoS. Hence, the CR is shown as
\begin{equation}
    C\left(\boldsymbol{\gamma},\boldsymbol{P}\right)=\sum_{n=1}^N\gamma_n\log_2\left(1+P_nSINR_{com,n}\right),
\end{equation}
where $P_n$ indicates the allocation of power to the $n$-th subcarrier, $SINR_{com,n}$ represents the SINR for UE communication on the $n$-th subcarrier, and $\gamma_n$ denotes subcarrier allocation factor, which can be expressed as follows
\begin{equation}
    \left.\gamma_n=\left\{\begin{matrix}1,&\text{ Case 1},\\0,&\text{ Case 2},\end{matrix}\right.\right.
\end{equation}
where case 1 represents the subcarrier $n$ is chosen for communication, and case 2 represents that is chosen for location sensing. For the radar sensing aspect, mutual information is used to evaluate sensing QoS. The mutual information of the sensed signal is shown as
\begin{equation}
    I\left(\boldsymbol{\gamma},\boldsymbol{P}\right)=\sum_{n=1}^{N}\left(1-\gamma_{n}\right)\log_{2}\left(1+P_{n}SINR_{rad,n}\right),
\end{equation}
where $SINR_{rad,n}$ indicates the SINR of radar echo signal on the $n$-th subcarrier.

Thus, the joint optimization problem formulates as
\begin{equation}
    \begin{aligned}
\min_{\boldsymbol{\gamma},\boldsymbol{P}}&\sum_{n=1}^NP_n & \\
\mbox{s.t.}\quad
&C1:C(\boldsymbol{\gamma},\boldsymbol{P})\geq{C_{min}},\forall n\in{N},  & \\
&C2:I\big(\boldsymbol{\gamma},\boldsymbol{P})\geq{I_{min}},\forall n\in N, &\\
&C3:0\leq P_n\leq P_{\max},\forall n\in N,&\\
&C4:\left.\gamma_n\in\{0,1\},\forall n\in N,\right.&
\end{aligned}
\end{equation}
where constraints $C1-C4$ respectively ensure communication QoS and sensing QoS, power constraints, and binary variables related to subcarrier allocation.

\subsection {Problem Solution}

In the formulation of optimization problem (41), it is observed that constraint $C4$ incorporates binary variables, while constraints $C1$ and $C2$ include two continuous variables related to the subcarrier allocation factor and power. Consequently, optimization problem (41) presents a non-convex optimization challenge. In order to solve this non-convexity, a methodology is introduced to solve this problem. Based on the solution approach for non-convex problems discussed in Section III, (41) is divided into two parts to tackle the non-convex challenge. Specifically, this decomposition yields sub-problems concerning subcarrier allocation and power allocation. These variables would be optimized sequentially.

For the subcarrier allocation sub-problem, the subcarrier allocation decision-making is based on the SINR for both communication and sensing. To achieve a demand-driven dynamic subcarrier allocation, a subcarrier allocation weighting $\eta$ is proposed. The mathematical representation is as follows
\begin{equation}
    \gamma_n=\begin{cases}1,&SINR_{com,n}\geq\eta SINR_{rad,n},\\0,&\mathrm{others}.\end{cases}
\end{equation}

The subcarrier allocation procedure is summarized in Algorithm 2. Once the subcarrier allocation is finalized, the original subcarrier set $N$ will be partitioned into two subsets: $N_{com}$ and $N_{rad}$. 

\begin{algorithm}
\caption{Subcarrier Dynamic Allocation}\label{alg:algorithm2}
\begin{algorithmic}[1]
\State \textbf{Initialization} the initial state of $i=1$, subcarrier set $N=1$, communication subcarrier subset $N_{com}=0$, location sensing subcarrier subset $N_{rad}=0$, and subcarrier allocation weighting $\eta$;
\State  \textbf{repeat} 
\State  \quad\textbf{for} $n=1$ to $N$ \textbf{do}
\State  \quad\quad \textbf{if} $SINR_{com,n}\geq\eta SINR_{rad,n}$
\State  \quad  \quad\quad $N^{i}_{com}\cup\{n\}\xrightarrow{}N^{i+1}_{com}$;
\State  \quad  \quad\quad $N^{i}\backslash\{n\}\xrightarrow{}N^{i+1}$;
\State  \quad \quad \textbf{else} 
\State  \quad  \quad \quad $N^{i}_{rad}\cup\{n\}\xrightarrow{}N^{i+1}_{rad}$;
\State  \quad  \quad \quad$N^{i}\backslash\{n\}\xrightarrow{}N^{i+1}$;
\State  \quad\quad \textbf{end} 
\State  \quad \textbf{end for} 
\State  \quad$i=i+1$;
\State  \textbf{until} $N=\emptyset$
\State  \textbf{Output} $N_{com},$ $N_{rad}.$
\end{algorithmic}
\end{algorithm}

Having determined the subsets of communication and location sensing subcarriers, we can further decompose optimization problem (41) into two sub-problems, denoted as (43) and (44). Sub-problem (43) concerns communication power allocation, while sub-problem (44) focuses on location sensing power allocation.

In sub-problem (43), the communication power allocation is optimized under the constraints of communication QoS. The optimization sub-problem is shown as
\begin{equation}
    \begin{aligned}\min_{\mathbf{P}}&\quad\sum_{c=1}^{N_{com}}P_{c}\\s.t.&\quad C1:C(\mathbf{P})\geq{C_{min}},\forall c\in N_{com},\\&\quad C2:0\leq P_c\leq P_{\max},\forall c\in N_{com},\end{aligned}
\end{equation}
where $C(\mathbf{P})=\sum_{c=1}^{N_{com}}\log_2\left(1+P_cSINR_{com,c}\right).$

In sub-problem (44), the location sensing power allocation  is optimized under the constraints of location sensing QoS. The optimization sub-problem is shown as \begin{equation}
\begin{aligned}\min_{\mathbf{P}}&\quad\sum_{s=1}^{N_{rad}}P_{s}\\s.t.&\quad C1:I\left(\mathbf{P}\right)\geq{I_{min}},\forall s\in N_{rad},\\&\quad C2:0\leq P_s\leq P_{\max},\forall s\in N_{rad},\end{aligned}
 \end{equation}
where $I\left(\mathbf{P}\right)=\sum_{s=1}^{N_{rad}}\log_2\left(1+P_s SINR_{rad,s}\right).$

Upon the decomposition of the original optimization challenge, it can be observed that both sub-problems (43) and (44) have transformed into convex optimization problems, with the primary variable being $\mathbf{P}$. To address these convex formulations, the well-established Lagrangian multiplier method is adopted. In particular, for the communication power allocation problem, the Lagrangian function corresponding to optimization problem (43) can be articulated as
\begin{equation}   \begin{aligned}&\mathcal{L}\left(P_{c},\lambda_{com},\mu_{com,1},\mu_{com,2}\right)\\
&=\sum_{c=1}^{N_{om}}P_c+\lambda_{com}\left[C_{min}-\sum_{c=1}^{N_{com}}\log_2\left(1+P_cSINR_{com,c}\right)\right]\\
    &+\mu_{com,1}\left(0-P_c\right)+\mu_{com,2}\left(P_c-P_{\max}\right).\end{aligned}
\end{equation}

To adhere to the communication QoS constraints and the Karush-Kuhn-Tucker (KKT) conditions of optimization problem, the minimization of communication subcarrier power allocation must satisfy the subsequent equation
\begin{equation}
    P_{c}^{\prime}=
    \begin{cases}
    P_{max},&{\iota _{com,c}}<\lambda_{com}^{\prime}-P_{max},\\
\lambda_{com}^{\prime}-{\iota _{com,c}},&\lambda_{com}^{\prime}-P_{max}<{\iota _{com,c}}<\lambda_{com}^{\prime},\\
0,&{\iota _{com,c}}>\lambda_{com}^{\prime},
    \end{cases}
\end{equation}
where ${\iota _{com,c}} = {1 \over {SIN{R_{com,c}}}}$, and $\lambda_{com}^{\prime}$ represents a constant of communication QoS constraints. The function is illustrated as
\begin{equation} \sum_{c=1}^{N_{com}}\log_2\left(1+P_{c}^{\prime}SINR_{com,c}\right)\geq{C_{min}}. 
\end{equation}

Analogously, to satisfy the location sensing QoS constraints and the KKT conditions of the optimization problem, the equation for minimizing the radar's location sensing subcarrier power allocation can be articulated as
\begin{equation}
    P_{s}^{\prime}=
    \begin{cases}
    P_{max},&{\iota _{rad,s}}<\lambda_{rad}^{\prime}-P_{max},\\
\lambda_{rad}^{\prime}-{\iota _{rad,s}},&\lambda_{rad}^{\prime}-P_{max}<{\iota _{rad,s}}<\lambda_{rad}^{\prime},\\
0,&{\iota _{rad,s}}>\lambda_{rad}^{\prime},
    \end{cases}
\end{equation}
where ${\iota _{rad,s}} = {1 \over {SIN{R_{rad,s}}}}$, and $\lambda_{rad}^{\prime}$ denotes a constant of sensing QoS constraints. The function is presented as
\begin{equation} \sum_{s=1}^{N_{rad}}\log_2\left(1+P_{s}^{\prime}SINR_{rad,s}\right)\geq{I_{min}}.
\end{equation}

Hence, Algorithm 3 outlines the specific procedure to solve problems (43) and (44).

\begin{algorithm}
\caption{Optimal Power Allocation for Communication and Sensing}\label{alg:algorithm3}
\begin{algorithmic}[1]
\State \textbf{Initialization} the maximum power $P_{max}$, communication QoS constraints $C_{min}$, sensing QoS constraints $I_{min}$, and communication and sensing subcarrier subset $N_{com}$ and $N_{rad}$;
\State  \textbf{for} $c=1$ to $N_{com}$ \textbf{do}
\State  \quad Calculate communication subcarrier power $P_{c}^{\prime}$ by (47);
\State  \quad Calculate communication Lagrange constants $\lambda_{com}^{\prime}$ by solving (46);
\State  \textbf{end for} 
\State  \textbf{for} $s=1$ to $N_{rad}$ \textbf{do}
\State  \quad Calculate sensing subcarrier power $P_{s}^{\prime}$ by (49);
\State  \quad Calculate sensing Lagrange constants $\lambda_{rad}^{\prime}$ by solving (48);
\State  \textbf{end for} 
\State  \textbf{Output} $P_{c}^{\prime},$  $P_{s}^{\prime}.$
\end{algorithmic}
\end{algorithm}

In summary, the processes for subcarrier and power allocation are presented in the 6G IRSLC network in Algorithm 4. Initially, the SINR of the UE's downlink communication and location echo signal is used as reference metrics, introducing a subcarrier allocation factor $\eta$, which could realize the dynamic allocation of subcarriers. Subsequently, under the constraints of QoS for SINR in the downlink communication and the sensing echo signal, based on the results of the dynamic allocation of subcarriers, power allocation is regulated flexibly based on the results of the dynamic allocation of subcarriers. This enables optimal allocation of power, ensuring the energy consumption of the system is minimized. 
 
\begin{algorithm}
\caption{6G IRSLC Network Energy Consumption Minimization}\label{alg:algorithm4}
\begin{algorithmic}[1]
\State  \textbf{Initialization} the initial state of $i=1$, subcarrier set $N=1$, communication subcarrier subset $N_{com}=0$, location sensing subcarrier subset $N_{rad}=0$, subcarrier allocation weighting $\eta$, the maximum power $P_{max}$, communication QoS constraints $C_{min}$, and sensing QoS constraints $I_{min}$;
\State  \textbf{for} $\eta$ = $\eta_{min}$ to $\eta_{max}$ \textbf{do}
\State  \quad Calculate $N_{com}$ and $N_{rad}$ by solving Algorithm 2;
\State  \quad Calculate $P_{c}^{\prime}$ and $P_{s}^{\prime}$ by solving Algorithm 3;
\State  \quad Calculate $\mathbf{P}(\eta)=P_{c}^{\prime}(\eta)+P_{s}^{\prime}(\eta)$;
\State  \textbf{end for}
\State  \textbf{Output} $\mathbf{P}=\arg\min\left\{\mathbf{P}(\eta)\right\},$ $\eta.$
\end{algorithmic}
\end{algorithm}

\section{SIMULATION RESULTS}
This section presents the analysis of UE location sensing and resource allocation outcomes in 6G IRSLC network under various parameters. In simulation, the number of carriers is adopted as $N=128$, with $N_c=10$ subcarriers within the relevant bandwidth and $M=10$ OFDM symbols. The millimeter-wave carrier frequency is $f_c=60$ GHz, and the system bandwidth is defined as $B=100$ MHz, where $B=1/T_s$. The speed of light is $c=3\times10^8$ m/s. For the simplicity in the analysis, the BS's transmit beamforming matrix is assumed to be $W_R$ and $W_C$ , which matches the transmitted signal $\mathbf{s}_m(n)$. Consequently, the equivalent signal is represented as $\mathbf{y}_{eq}\in\mathbb{C}^{N_{T}\times1}$, and its detailed expression is as follows
\begin{equation}
    y_{eq}(n)=\frac{1}{\sqrt{M}}e^{-\mathrm{j}\pi(t-1)},t\in{N_T}.
\end{equation}

Moreover, during the location sensing phase, the BS receives $K=2$ echo signals. It is also assumed that the number of sensed echo signals with interference received by the UE is the same. UE is randomly distributed within a sensing communication radius of $50$ m, and the positions of buildings are also randomly distributed within this area. The relative velocity of the UE is denoted by $v_{UE}\in[0,20]$ m/s, and the Doppler shift due to this relative motion is represented by $f_d\in[0,8]$ kHz. The transmitting angle of the ULA is set as $\phi\in[-\pi,\pi]$ Rad, and $\alpha_k={e^{\mathrm{j}\omega_k}}/{(\tau_kc)^2}$, where $\omega_k\in[-\pi,\pi)$ and $\tau_kc=\|\mathbf{Q}_\mathrm{BS}-\mathbf{Q}_\mathrm{UE}\|_2+\|\mathbf{Q}_\mathrm{UE}-\mathbf{Q}_k\|_2+\|\mathbf{Q}_\mathrm{BS}-\mathbf{Q}_k\|_2$ represent the round-trip propagation distance from the BS to the UE, and $\mathbf{Q}_k$ stands for the coordinates of buildings randomly positioned within the area. In the communication phase, parameters of the communication channel model include: reference distance $d_0=1$ m, path loss exponent $\varepsilon=2.9$, and log-normal shadow fading $\sigma_{\zeta}^2=5.7$ dB. In the subcarrier allocation, the maximum power assigned to the $n$-th subcarrier is set at $P_{max}=50$ W. The noise power for radar location sensing and communication is denoted by $\sigma_R^2=1\times10^{-14}$ W/Hz and $\sigma_C^2=1\times10^{-14}$ W/Hz respectively. The QoS constraints for radar position sensing and communication are set at $I_{min}=600$ and $C_{min}=200$, respectively.

\subsection {UE Loaction Sensing}

The estimated AoA is derived from Eq. (37) in Section III, as shown in Fig. 2. Fig. 2 shows the estimated AoA value after 100 Monte Carlo simulations. In this simulation, the launch angle is designed as $\phi=20^\circ$, and transmission antenna and receive antenna is set as $N_T=N_R=8$. According to the channel reciprocity principle, the theoretical value of AoA should also be 20 degrees. It can be seen that the angle value estimated by the location sensing algorithm based on IRSLC is in good agreement with the theoretical value. This demonstrates that the location sensing estimation algorithm proposed in this paper has good angle estimation performance in the 6G multiple cells IRSLC system. In addition, as the AoA of IDR is not the focus of this article, the IDR angles in the remaining two directions received by the BS are not listed.

\begin{figure}[H]
    \centering
    \subfigure[AoA estimation results by IRSLC-based location estimate algorithm]{
    \includegraphics[width=0.6\linewidth]{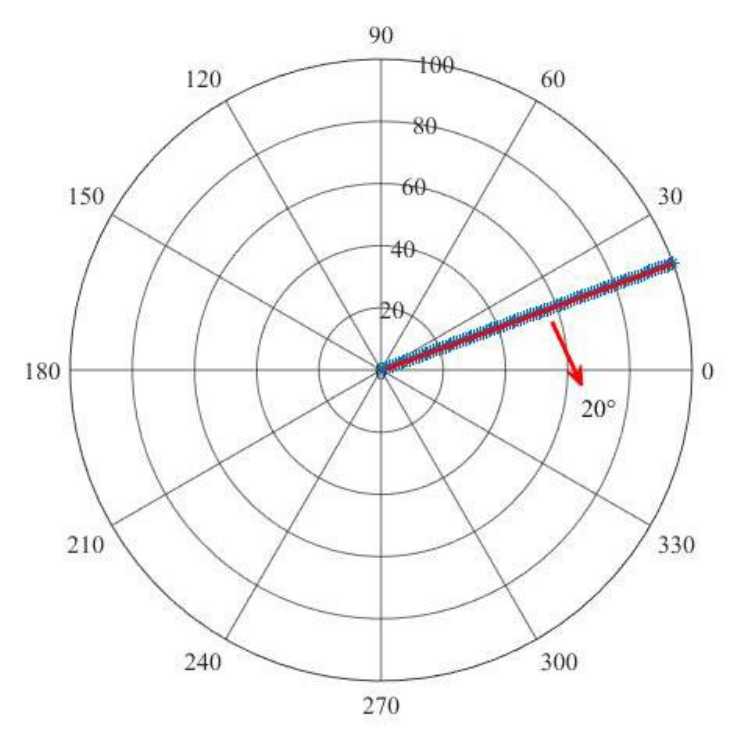}} 
    \quad
    \subfigure[Specific parts of AoA estimation]{
    \includegraphics[width=0.75\linewidth]{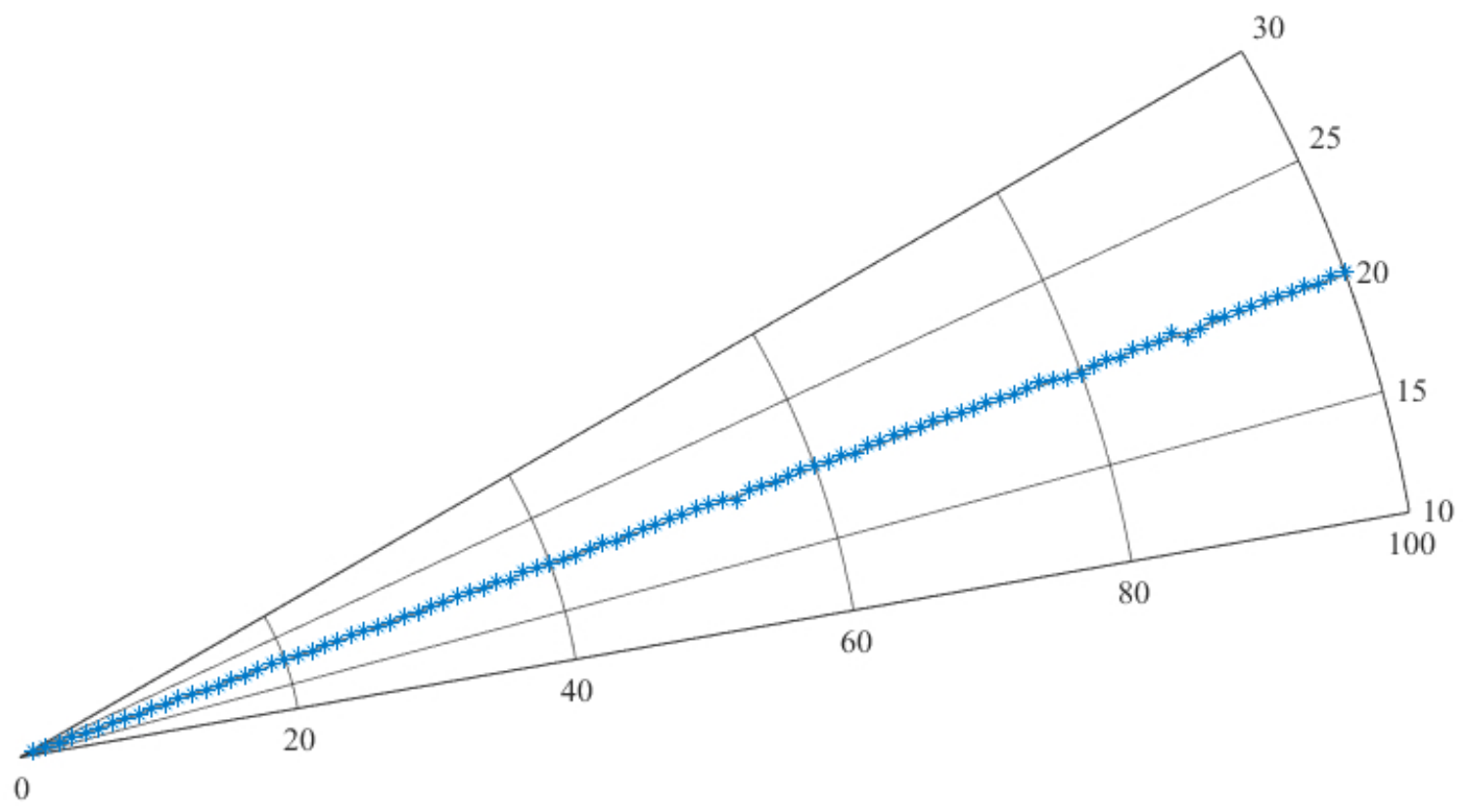}} 
    \caption{AoA estimation by IRSLC-based location estimate algorithm with Monte Carlo.
}
    \label{t一}
\end{figure}

In Fig. 3, the cumulative distribution function of location sensing errors of UE, which is presented under various algorithms and 100 Monte Carlo simulations, where the total numbers of antennas are $N_T+N_R=8$ and $N_T+N_R=16$. The proposed method exhibits remarkable superiority in error distribution. When the antennas are set to $16$, the majority of errors remaining within the $(0.05,0.2)$ interval. And the antennas are set to $8$, the majority of errors remaining within the $(0.1,0.2)$ interval. Obviously, as the quantity of antennas grows, the coverage range and sampling density of the system to the environment are improved, leading to an improvement in sensing accuracy and a gradual reduction in location error. Regarding the classical location sensing algorithms, TLS-ESPRIT and multiple signal classification (MUSIC), both exhibit errors within the small range of $(0.1,0.2)$, which are typically considered acceptable. However, when the total number of antennas is $16$, TLS-ESPRIT and MUSIC algorithms also demonstrate considerable errors distributed in the wider range of $(0.3,0.7)$, which are often deemed unsatisfactory in simulations. In addition, when the antennas are set to $8$, the location error presented is mostly distributed outside the $0.7$ range, which is difficult to accept in simulation. Therefore, the proposed algorithm can exhibit significant advantages under different total antenna numbers, and its advantages are more pronounced in small-scale antennas.

\begin{figure}
    \centering
    \includegraphics[width=0.8\linewidth]{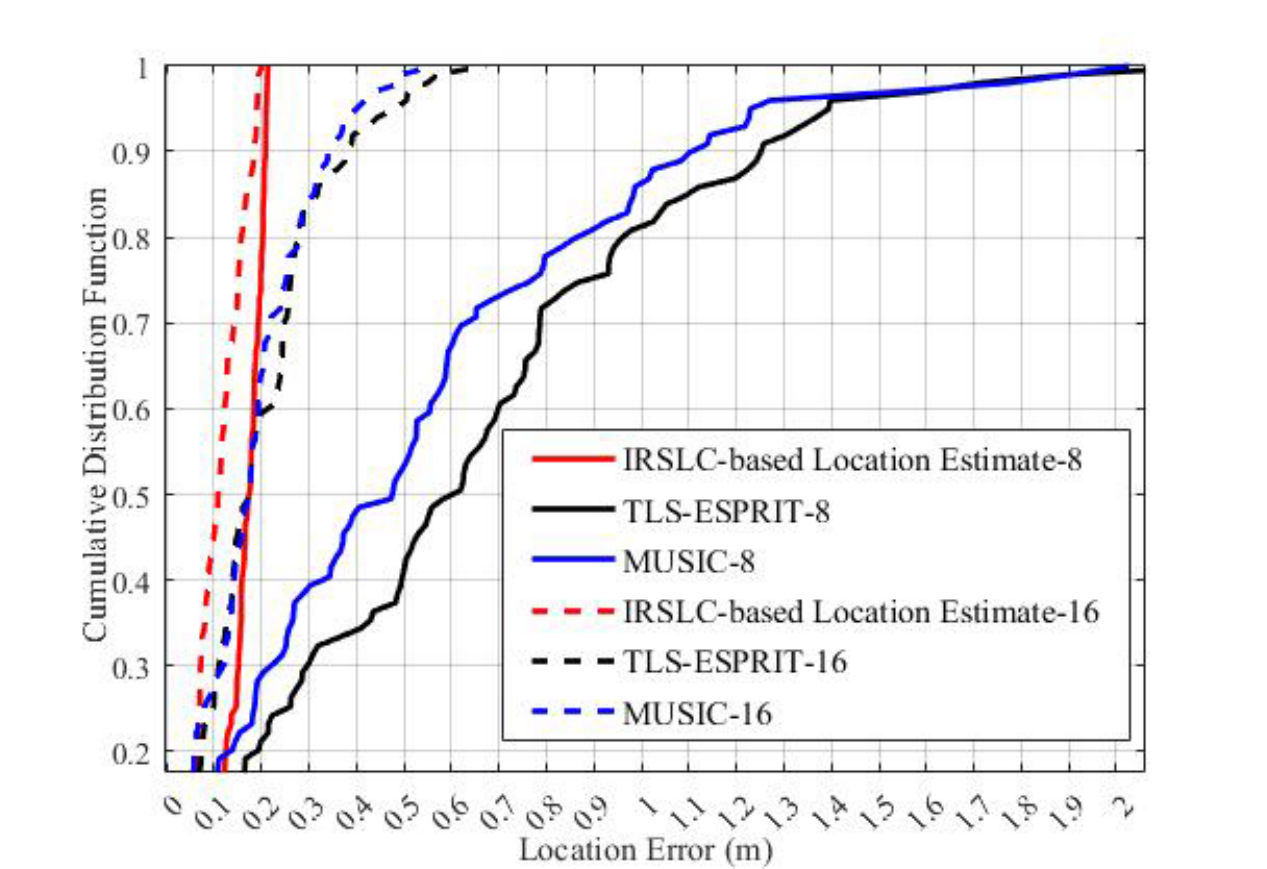}
    \caption{Cumulative distribution of location sensing error.}
    \label{fig:enter-labe2}
\end{figure}

\subsection {Joint Subcarrier and Power Allocation Results}
Fig. 4 depicts the subcarrier allocation results under different subcarrier allocation weights $\eta$, with Figs. 4-6 all conducted with a total antenna count of 16 (i.e. $N_T=N_R=8$). As observed, with the continuous variation of subcarrier allocation weight $\eta$, both communication subcarriers and radar location sensing subcarriers are dynamically adjusted. In the location sensing phase, a significant amount of power is allocated to radar location sensing. Therefore, to ensure minimal system energy consumption while meeting the QoS requirements of radar sensing, the number of allocated subcarriers tends to decrease. As the UE location is captured, the subcarrier allocation undergoes changes. To ensure minimal system energy consumption while satisfying communication QoS, the number of communication subcarriers allocated gradually decreases, while the power allocation for communication gradually increases.

\begin{figure}
    \centering
    \includegraphics[width=0.8\linewidth]{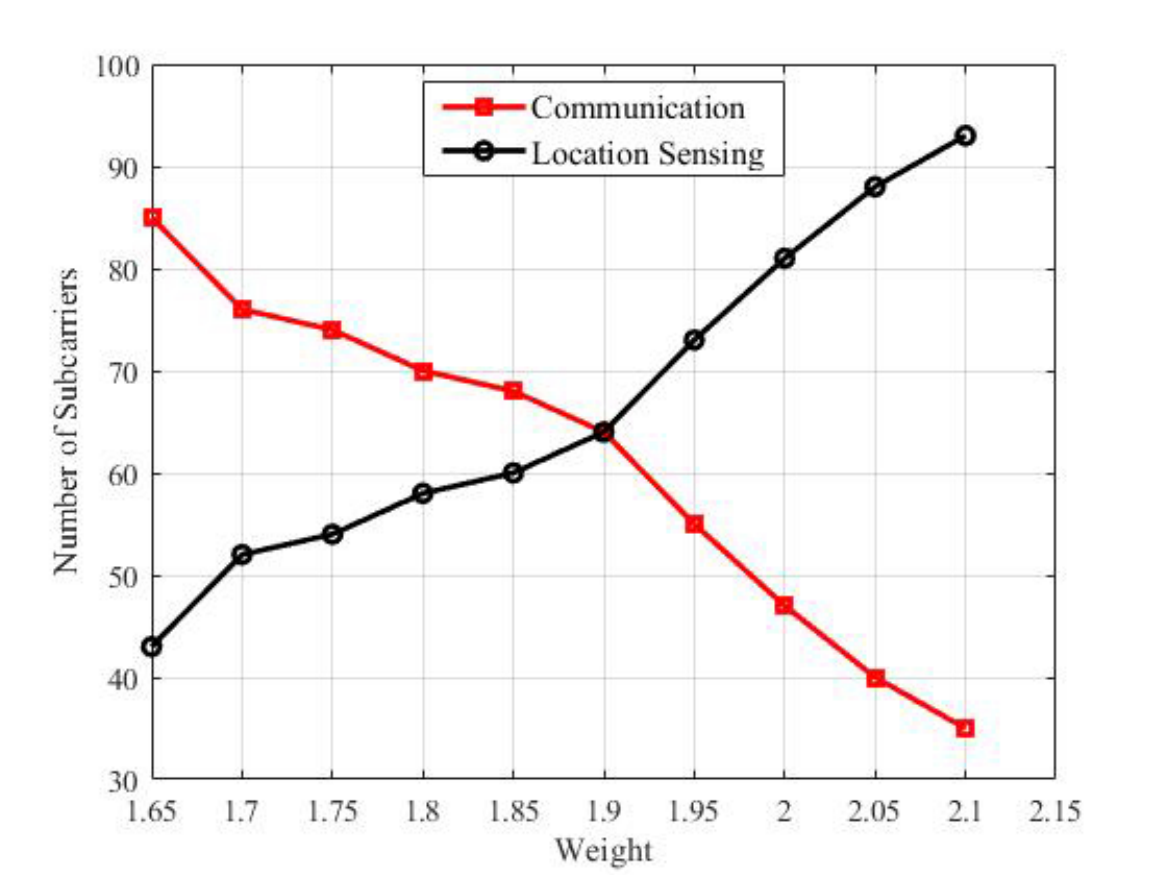}
    \caption{Subcarrier allocation results with different subcarrier allocation weights $\eta$.}
    \label{fig:enter-labe3}
\end{figure}

Fig. 5 showcases the power allocation results for communication and radar sensing under different subcarrier allocation weights $\eta$. One can observes that as the weight $\eta$ growth, the power allocation of communication continuously rises, given that the communication phase primarily aims to fulfill UE communication quality and demands. Meanwhile, the power for radar sensing decreases consistently since, during the communication phase, radar sensing is ongoing but isn't the primary focus.

\begin{figure}
    \centering
    \includegraphics[width=0.8\linewidth]{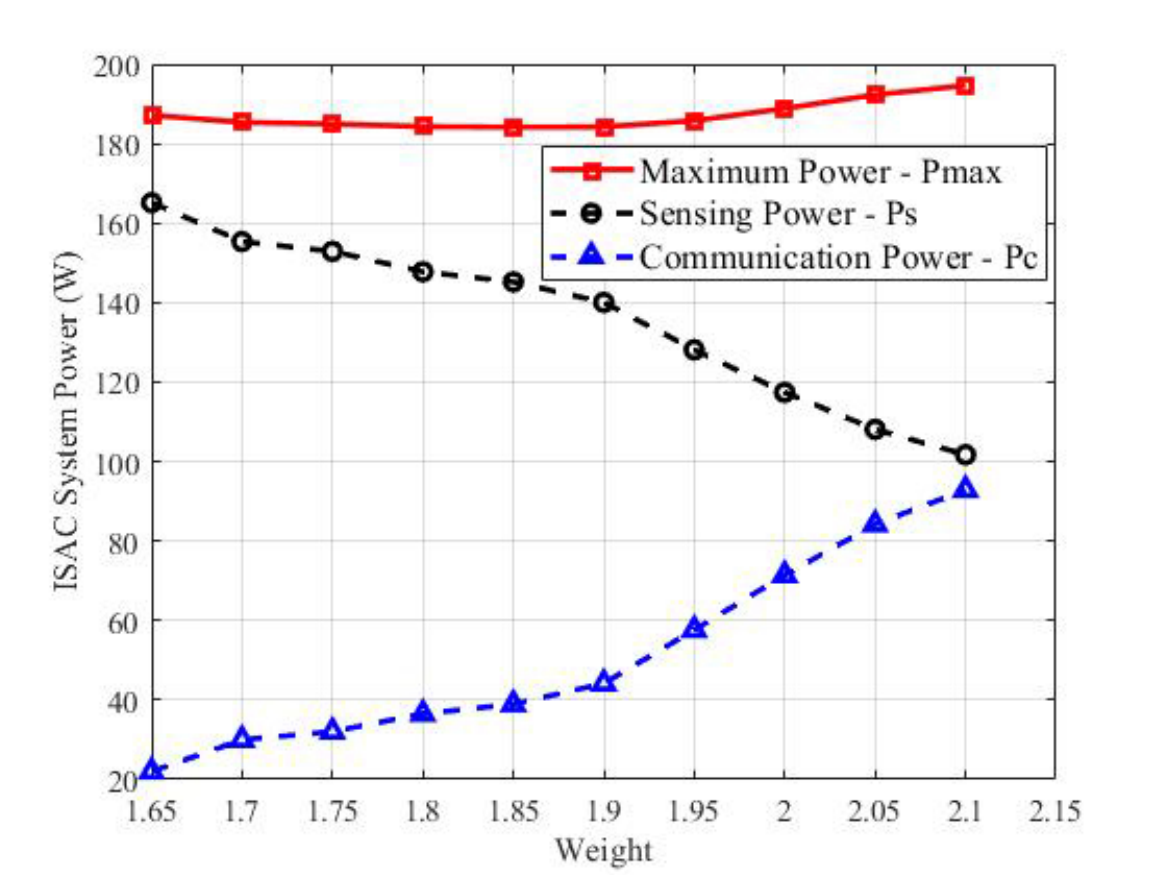}
    \caption{Power allocation results with different subcarrier allocation weights $\eta$.}
    \label{fig:enter-labe4}
\end{figure}

Subsequently, we consider the case of the lowest system energy consumption in Fig. 5. (i.e. $\eta=1.9$). In this context, Fig. 6 illustrates the results for radar location sensing and communication after joint resource allocation, where the number of subcarriers set is 128. Observing the allocation outcomes, it becomes apparent that when subcarriers are designated for location sensing, the subcarriers cannot be allocated by the communication side at this time. This corresponds to the allocation results of the first 6 subcarriers in Fig. 6(a) and Fig. 6(b). In contrast, when subcarriers are allocated to communication, corresponding subcarriers cannot be allocated by the location sensing, which corresponds to the channel index blank part of the location sensing allocation result. In addition, in order to better obtain the UE's location status information, the location sensing power is usually higher than the communication power. From Figs. 5 and 6, with an allocation weight of $\eta=1.9$, the system can achieve the minimum energy consumption under this joint allocation.

\begin{figure}
\renewcommand{\thefigure}{6}
    \centering
    \subfigure[Sensing location power under optimal allocation $\eta=1.9$]{
    \includegraphics[width=0.8\linewidth]{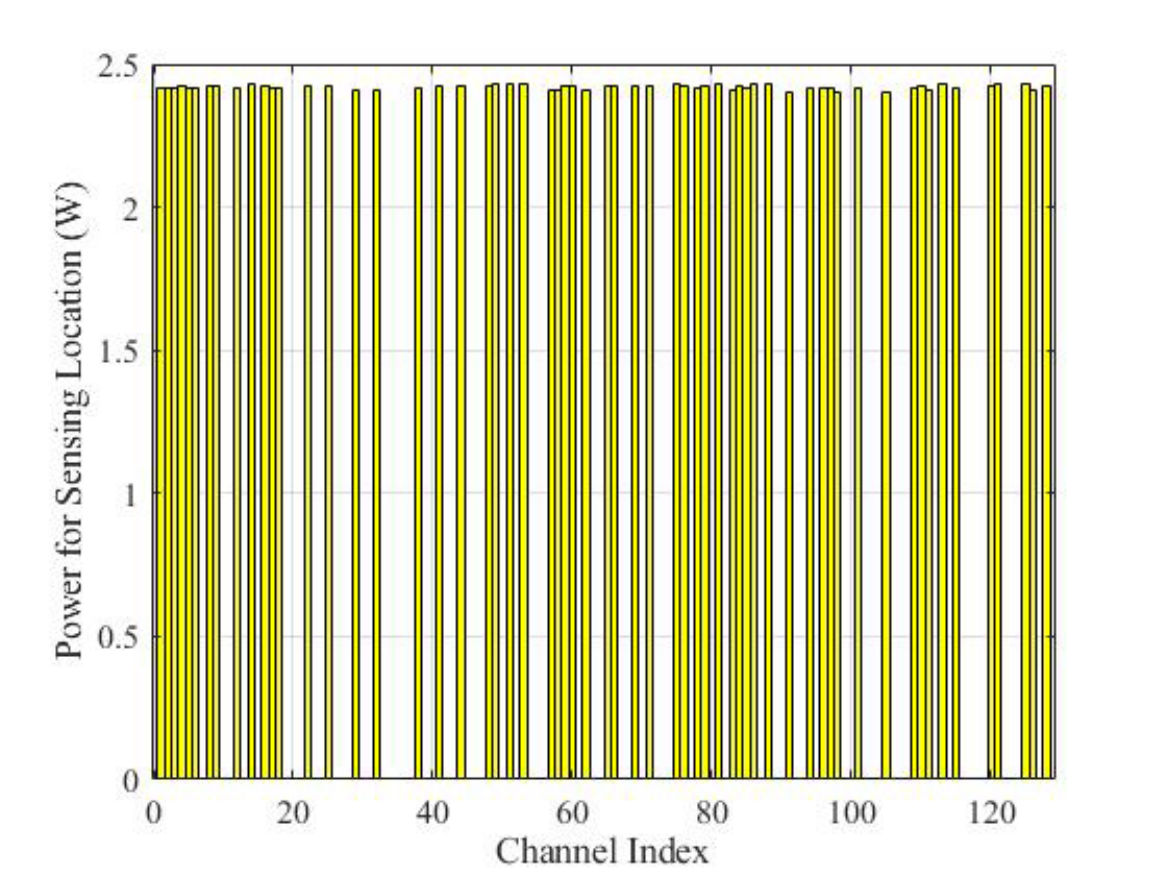}} 
    \quad
    \subfigure[Communication power under optimal allocation $\eta=1.9$]{
    \includegraphics[width=0.8\linewidth]{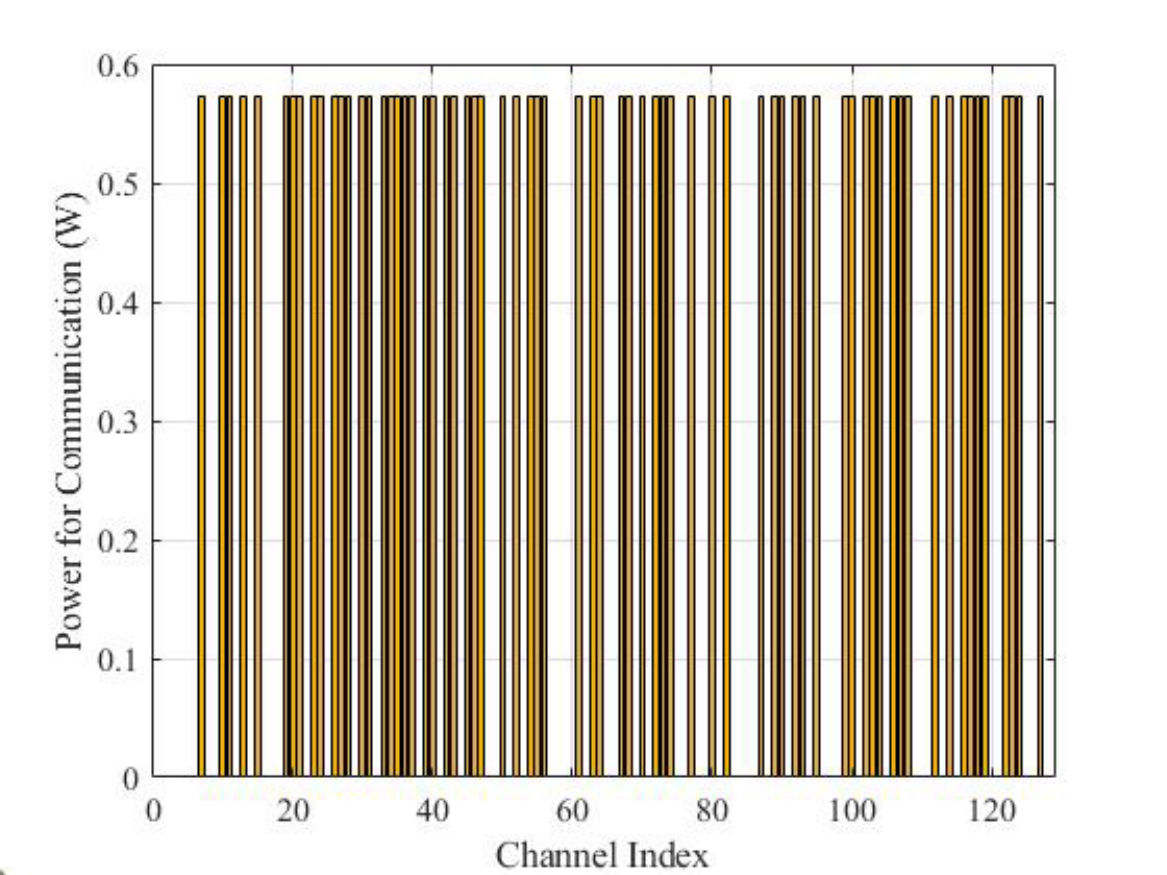}} 
    \caption{Joint subcarrier and power allocation results for radar location sensing and communication.
}
\end{figure}

To reveal the advantage of the suggested algorithm, in Fig. 7, the Joint Subcarrier and Power Allocation algorithm according to IRSLC (IRSLC-JSPA) is compared with the following three algorithms. For Resource Allocation Comparison Algorithm 1 (RACA1), the subcarriers are allocated according to Algorithm 2 proposed in this paper, and the power for communication and radar sensing is uniformly distributed. For Resource Allocation Comparison Algorithm 2 (RACA2), both communication and radar sensing subcarriers are randomly allocated, but the power distribution is based on Algorithm 3 presented in this paper. For Resource Allocation Comparison Algorithm 3 (RACA3), the communication and radar sensing subcarriers are randomly assigned, and power of communication and location sensing is uniformly distributed. As shown in Fig. 7, the proposed IRSLC-JSPA algorithm consumes the least power among all algorithms while simultaneously meeting the minimum performance requirements for communication and radar location sensing. This confirms the efficacy of the proposed scheme. Furthermore, from the comparison, it is evident that when the subcarriers are randomly allocated, there is a noticeable increase in power consumption. This indicates that the subcarrier allocation method proposed in this study can dynamically allocate subcarriers based on system demands, thereby significantly reducing system energy consumption.

 \begin{figure}
     \centering
     \includegraphics[width=0.8\linewidth]{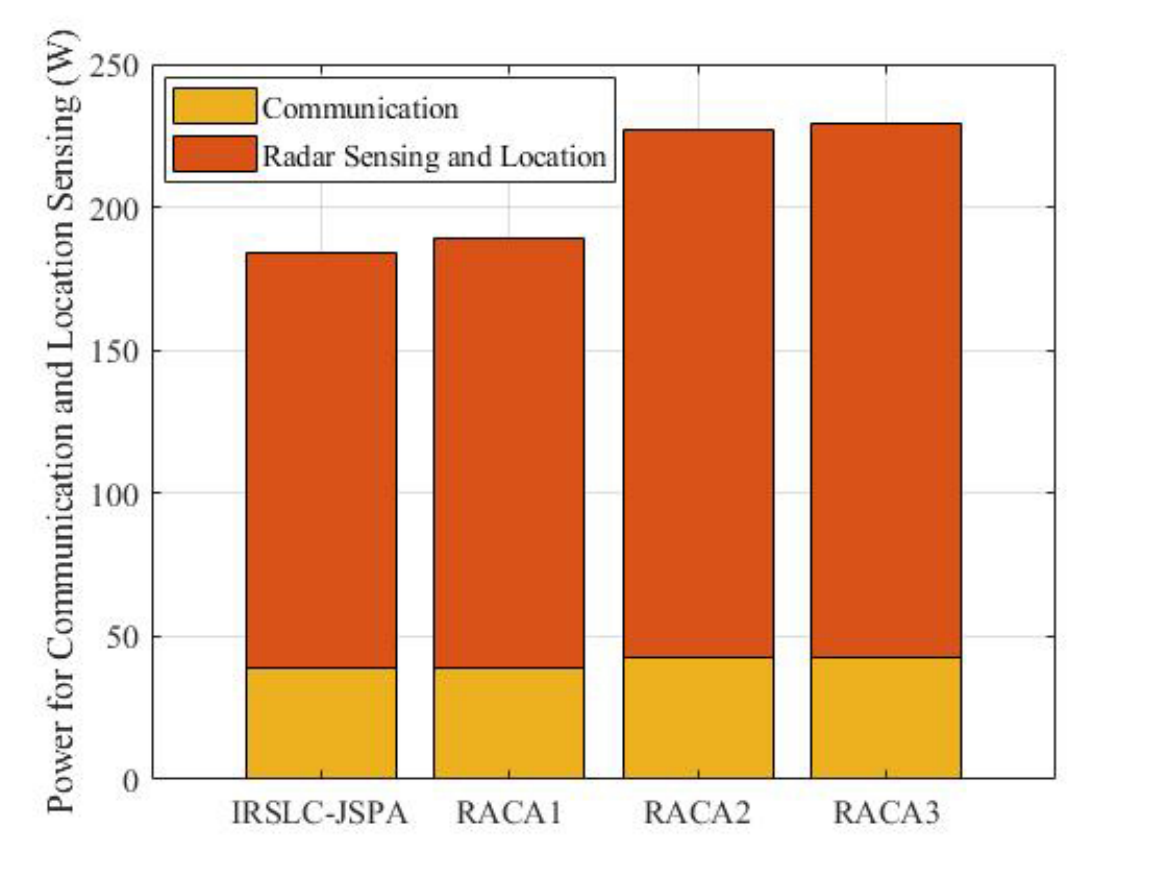}
     \caption{Comparison algorithm for subcarrier assignment and power allocation.}
     \label{fig:enter-labe5}
 \end{figure}

\section{CONCLUSION}
In this paper, a joint location sensing and resource allocation scheme within the IRSLC system is designed. Initially, an IRSLC system is established that uses IRSLC BS to emit radar sensing signals to acquire UE location.  This information is then utilized for allocating the joint subcarriers and power, which can reduce the system energy consumption. Subsequently, a novel location sensing scheme is introduced, which is capable of effectively handling challenges such as multipath interference, random channel fading, and Doppler effects, leading to a notable enhancement in positioning accuracy. Ultimately, a joint subcarrier and power allocation scheme based on UE location information and radar sensing SINR is designed, which aids the IRSLC system in effectively reducing energy consumption and achieving energy-saving objectives. Overall, this is an energy-saving solution that simultaneously performs location sensing and resource allocation, which is of great significance for saving system energy and achieving green communication.

%Appendix one text goes here %\cite{Roberg2010}.

%\section{}

\section*{Appendix A}

According to Eqs. (3) and (32), the gradient calculation formula for $\mathbf{Q}_{UE}$ can be obtained by
\begin{equation}
\begin{aligned}
&\xi_{r,n,m}^{\mathbf{Q}_{UE,2}}\left(\mathbf{\hat{U}}_{[i]}\right)=-j2\pi f_{n}^{\prime}d_{A}\left(\frac{\mathbf{e}_{X}}{\left\|\mathbf{\hat{Q}}_{UE,[i]}-\mathbf{Q}_{BS}\right\|_{2}}-\right.\\&\left.\frac{\left(\hat{\mathbf{Q}}_{UE,[i]}-\mathbf{Q}_{BS}\right)^T\mathbf{e}_X\left(\hat{\mathbf{Q}}_{UE,[i]}-\mathbf{Q}_{BS}\right)}{\left\|\hat{\mathbf{Q}}_{UE,[i]}-\mathbf{Q}_{BS}\right\|_2^3}\right)\chi\left(\hat{\mathbf{Q}}_{UE,[i]}\right),
\end{aligned}
\end{equation}
\begin{equation}
    \chi\left(\hat{\mathbf{Q}}_{UE,[i]}\right)=\left(\iota\odot\boldsymbol{\alpha}_{r,n}^\mathrm{H}(\hat{\mathbf{Q}}_{UE,[i]})\right)\mathbf{W}_R\mathbf{s}_m(n),
\end{equation}
where $\imath=[0,1,\cdots,N_\mathrm{T}-1]\in\mathbb{R}^{N_\mathrm{T}\times1}$.

According to Eqs. (3), (4), (31) and (52), the gradient calculation formula for $\mathbf{Q}_{UE}$ can be obtained by
\begin{equation}
\begin{aligned} 
&\xi_{r,n,m}^{\mathbf{Q}_{UE,1}}\left(\mathbf{\hat{U}}_{[i]}\right)=\left[-\mathrm{j}4\pi(f_{n}-f_{d})\frac{\hat{\mathbf{Q}}_{UE,[i]}-\mathbf{Q}_{BS}}{c\left\|\hat{\mathbf{Q}}_{UE,[i]}-\mathbf{Q}_{BS}\right\|_{2}}\right.
\\&-j2\pi f_{n}^{\prime}d_{A}\left(\frac{\mathbf{e}_{X}}{\left\|\hat{\mathbf{Q}}_{UE,[i]}-\mathbf{Q}_{BS}\right\|_{2}}-\left(\hat{\mathbf{Q}}_{UE,[i]}-\mathbf{Q}_{BS}\right)^{T}\bullet\right.
\\&\left.\left.\frac{\mathbf{e}_{X}\left(\hat{\mathbf{Q}}_{UE,[i]}-\mathbf{Q}_{BS}\right)}{\left\|\hat{\mathbf{Q}}_{UE,[i]}-\mathbf{Q}_{BS}\right\|_{2}^{3}}\right)\Big(1+\chi\left(\hat{\mathbf{Q}}_{UE,[i]}\right)\Big)\right]
\\&\bullet\left(\mathsf{e}^{-\mathrm{j}2\pi(f_n-f_d)\hat{\tau}_{0,[i]}}b_{r,n}(\mathbf{\hat{Q}}_{UE,[i]})\boldsymbol{\alpha}_{n}^{\mathrm{H}}(\mathbf{\hat{Q}}_{UE,[i]})\mathbf{W}_R\mathbf{s}_m(n)\right).
\end{aligned}
\end{equation}

\ifCLASSOPTIONcaptionsoff
  \newpage
\fi

\bibliographystyle{IEEEtran}
\bibliography{Reference}
% ==== SWITCH OFF the BIO for submission
% ==== SWITCH OFF the BIO for submission
\vfill

\begin{IEEEbiography}[{\includegraphics[width=1in,height=1.25in,clip,keepaspectratio]{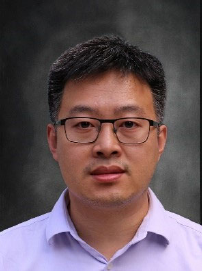}}]{Haijun Zhang} (Fellow, IEEE) is currently a Full Professor and Dean in School of Computer and Communications Engineering at University of Science and Technology Beijing, China. He was a Postdoctoral Research Fellow in Department of Electrical and Computer Engineering, the University of British Columbia (UBC), Canada. He serves/served as Track Co-Chair of VTC Fall 2022 and WCNC 2020/2021, Symposium Chair of Globecom’19, TPC Co-Chair of INFOCOM 2018 Workshop on Integrating Edge Computing, Caching, and Offloading in Next Generation Networks, and General Co-Chair of GameNets’16. He serves/served as an Editor of IEEE Transactions on Wireless Communications, IEEE Transactions on Information Forensics and Security, and IEEE Transactions on Communications. He received the IEEE CSIM Technical Committee Best Journal Paper Award in 2018, IEEE ComSoc Young Author Best Paper Award in 2017, IEEE ComSoc Asia-Pacific Best Young Researcher Award in 2019. He is a Distinguished Lecturer of IEEE and IEEE Fellow.
\end{IEEEbiography}

\begin{IEEEbiography}[{\includegraphics[width=1in,height=1.25in,clip,keepaspectratio]{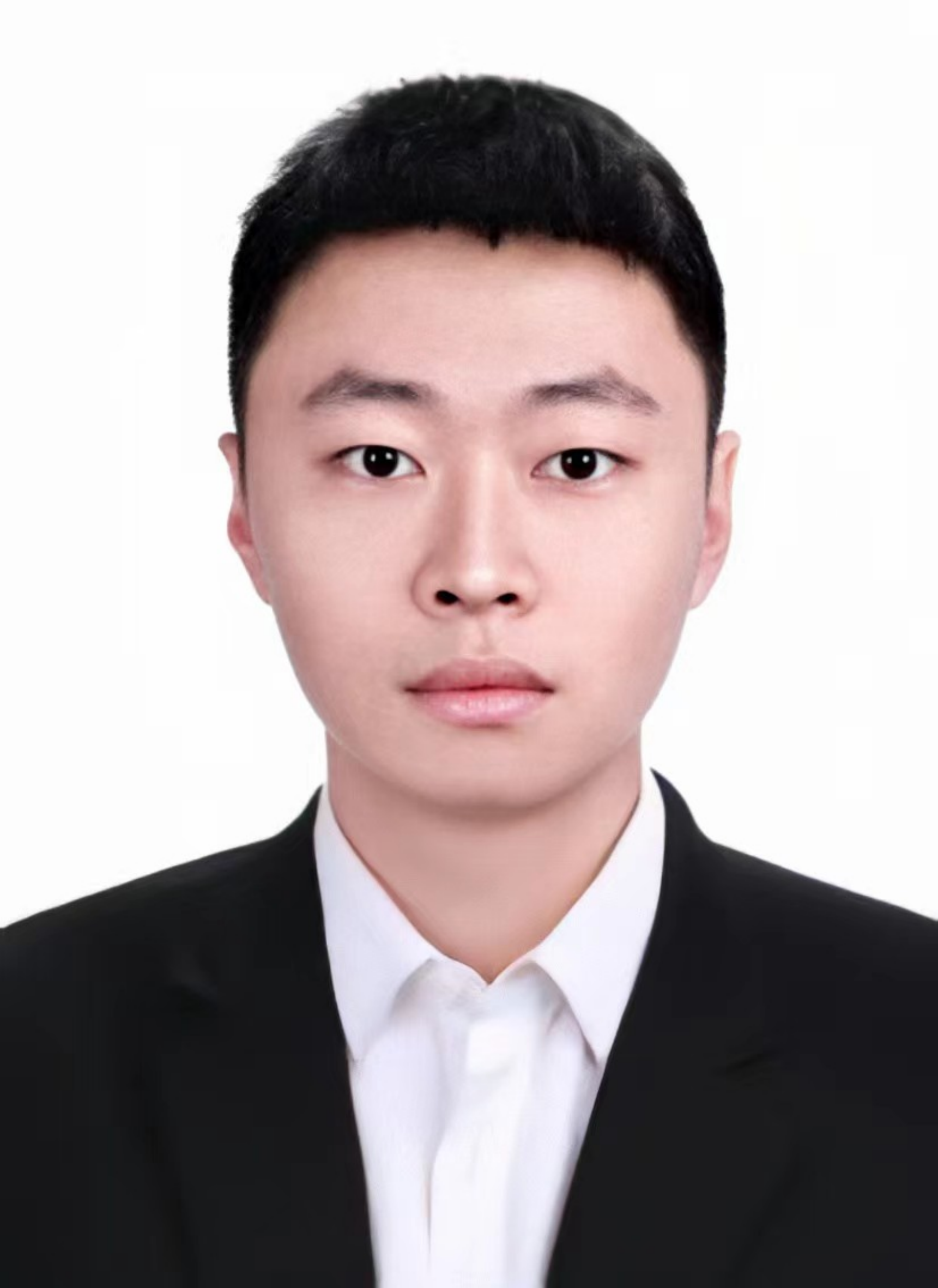}}]{Bowen Chen} received the B.S. degree from the School of Computer and Communication Engineering, University of Science and Technology of Beijing, Beijing, China, in 2021. He is currently pursuing the M.S. degree at University of Science and Technology Beijing, China. His research interests include resource allocation in 6G wireless communication and location sensing.
\end{IEEEbiography}

\begin{IEEEbiography}[{\includegraphics[width=1in,height=1.25in,clip,keepaspectratio]{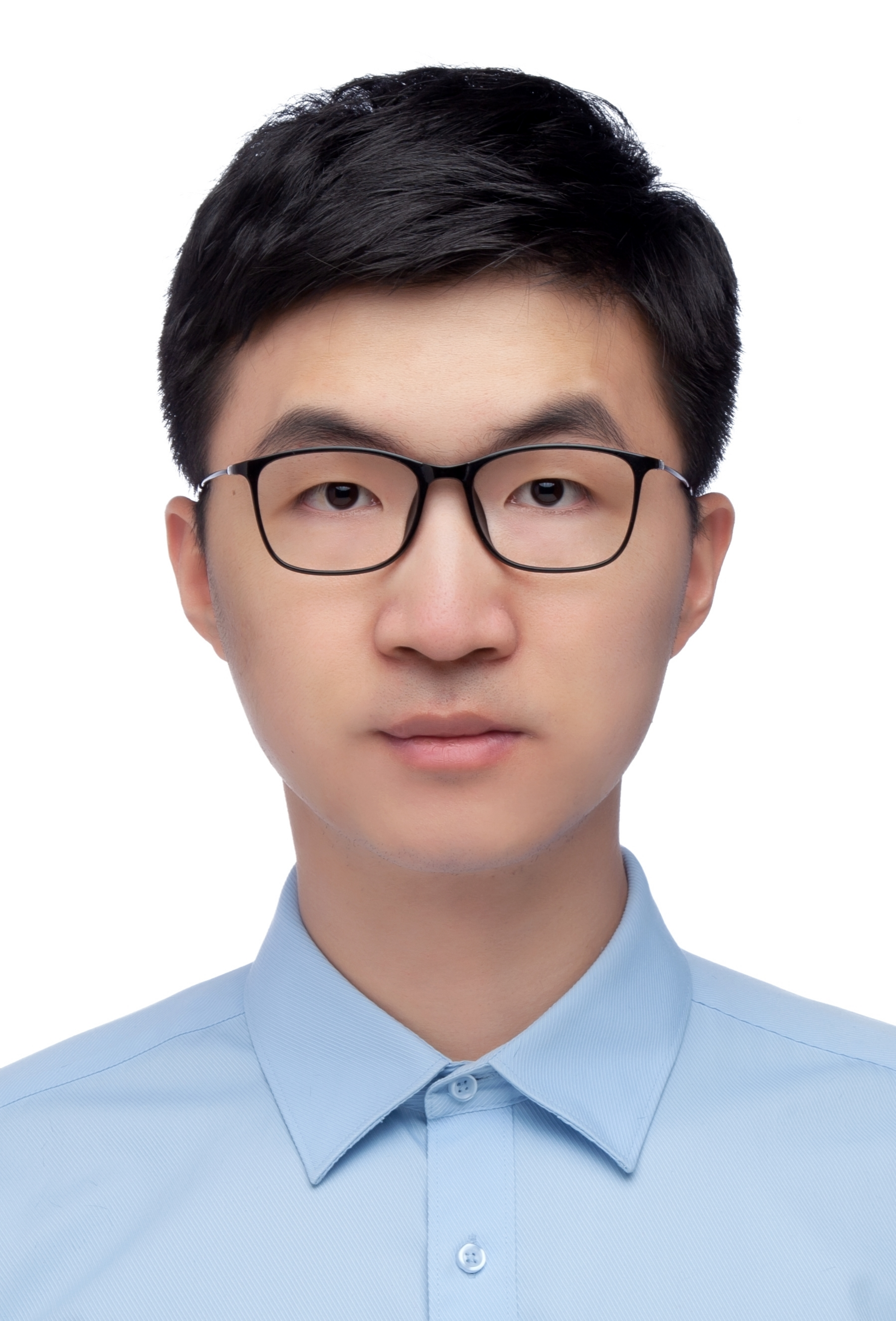}}]{Xiangnan Liu} received the B.S. degree from the School of Computer and Communication Engineering, University of Science and Technology of Beijing, Beijing, China, in 2019. He is currently pursuing his Ph.D. degree at University of Science and Technology Beijing, China. His research interests include access control, beamforming, and resource allocation in 6G wireless communication. He serves as a reviewer for IEEE TRANSACTIONS ON WIRELESS COMMUNICATIONS, IEEE NETWORK, and several journals.
\end{IEEEbiography}

\begin{IEEEbiography}[{\includegraphics[width=1in,height=1.25in,clip,keepaspectratio]{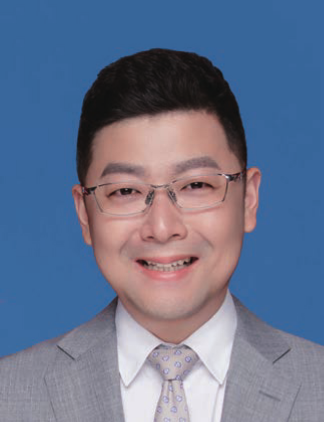}}]{Chao Ren} received the B.Eng. degree from the Ocean University of China, Qingdao, China, in 2011, and the Ph.D. degree from Xidian University, Xi’an, China, in 2017. From 2015 to 2017, he was with the University of Alberta, Edmonton, AB, Canada, as a Joint Ph.D. Student sponsored by the China Scholarship Council (CSC). He is currently with the University of Science and Technology Beijing, China. His current research interests include  cooperative communication technology in advanced wireless networks and its new derivative solutions. He is an Editor of Frontiers in Computer Science and also a co-chair of UCOM WKSP 2023-2024.
\end{IEEEbiography}


\end{document}